\documentclass[nofootinbib,superscriptaddress,floatfix]{revtex4}

\usepackage{amsmath,amssymb,bbm}
\usepackage{epsfig}

\newcommand{\as}{\bar\alpha_s}
\newcommand{\gc}{\gamma_0}
\newcommand{\xc}{\chi(\gamma_0)}
\newcommand{\xggc}{\chi''(\gamma_0)}
\newcommand{\Qs}{Q_s}
\newcommand{\Q}{Q_s^2(Y)}

\def\wid{0.7\columnwidth}

\begin{document}

\begin{flushright}
{\bf LBNL-57569}\\
{\bf CPHT-RR-026.0505}\\
{\bf DFF-425/05/05}
\end{flushright}

\title{The high energy asymptotics of scattering processes in QCD
}

\date\today

\author{R. Enberg}

\affiliation{Theoretical Physics Group,
Ernest Orlando Lawrence Berkeley National Laboratory,
CA 94720, USA}

\affiliation{Centre de Physique Th{\'e}orique, Unit\'e mixte de
recherche du CNRS (UMR 7644),
{\'E}cole Polytechnique, 91128~Palaiseau, France}

\author{K. Golec-Biernat}

\affiliation{Institute of Nuclear Physics, Polish Academy of Sciences,
Cracow, Poland}

\author{ S. Munier}

\affiliation{Centre de Physique Th{\'e}orique, Unit\'e mixte de
recherche du CNRS (UMR 7644),
{\'E}cole Polytechnique, 91128~Palaiseau, France}

\affiliation{Dipartimento di Fisica, Universit\`a di Firenze, via Sansone 1,
50019~Sesto F., Florence, Italy.}

\begin{abstract}
High energy scattering in the QCD parton model
was recently shown to be a
reaction-diffusion process and, thus, to lie in the
universality class of the stochastic Fisher-Kolmogorov-Petrovsky-Piscounov
equation. We recall that the latter appears naturally in the context of
the parton model.
We provide a thorough numerical analysis of the mean field approximation,
given in QCD by the Balitsky-Kovchegov equation.
In the framework of a simple stochastic 
toy model that captures the relevant features
of QCD, we discuss and illustrate the universal properties of
such stochastic models.
We investigate, in particular, the validity of 
the mean field approximation and how it is broken by fluctuations.
 We find that the mean field approximation is a good approximation in 
the initial stages of the evolution in rapidity.
\end{abstract}

\maketitle


\section{Introduction}

The description of the rise of hadronic cross sections with the center of mass
energy $\sqrt{s}$ within quantum chromodynamics (QCD)
has been a challenging issue. 
Old experimental results, 
e.g.\ on proton scattering,
do not yet have a satisfactory theoretical explanation.
So far, no realization has been found for
the Froissart bound, which is a consequence of the unitarity of the $S$-matrix
and which is a bound on the high energy behavior of the total cross sections,
$\sigma\leq (1/m_\pi^2) \ln^2 s$.

A perturbative calculation of the evolution of scattering amplitudes 
with energy, achieved by Balitsky,
Fadin, Kuraev and Lipatov (BFKL) 30 years ago \cite{BFKL}
and subsequently improved to next-to-leading order
\cite{NLBFKL}, is apparently successful in describing the
rise of photon-proton cross sections when the virtuality of
the photon is high enough.
However, the BFKL equation, being linear, is not compatible
with the Froissart bound when it is extrapolated to very high energies.
Complying with the limits set by unitarity requires the introduction of
nonlinear terms in the
evolution equations. A first step in this direction was
taken by Gribov, Levin and Ryskin (GLR) in 1981 \cite{GLR}, 
and by Mueller and Qiu (MQ) in 1986 \cite{MQ}.
Subsequently, more involved QCD evolution
equations were derived within different
formalisms. Balitsky, Jalilian-Marian, Iancu, McLerran,
Weigert, Leonidov and Kovner (B-JIMWLK) \cite{B,JIMWLK} 
have developed a comprehensive 
approach to high energy scattering from different but converging points
of view.
Technically, they were able to write the high energy evolution
of QCD amplitudes in the form
of a functional integro-differential equation, that can also be expressed
as a Langevin equation or as an infinite hierarchy of coupled
differential equations.
A much simpler equation was obtained by Balitsky \cite{B} 
and by Kovchegov \cite{K} (BK)
in the particular physical context where the target is a large nucleus.
The latter turns out to be a specific limit of the former, and its
structure is similar to the GLR-MQ equation.

Meanwhile, on the phenomenological side, 
Golec-Biernat and W\"usthoff showed \cite{GBW}
that unitarization effects, that cannot be taken into account by a linear
evolution such as the BFKL equation,
may have already be seen at the DESY electron-proton collider HERA.
Their observation gave a new impetus to the field and triggered many
phenomenological and theoretical studies.
Furthermore, the model they had proposed
pointed towards a new scaling, ``geometric scaling'', which was subsequently
found in the HERA data \cite{SGK}.

Much insight and intuition 
has been gained from numerical studies: even before 
the B-JIMWLK formalism had been developed,
Salam performed numerical calculations of scattering amplitudes
near the unitarity limits in the framework of a Monte-Carlo
implementation \cite{Salam} of Mueller's color dipole model \cite{CD}.
More recently, much effort has been devoted to 
obtaining numerical solutions of the BK and B-JIMWLK equations
\cite{AB,L2001,GMS,AAMSW,RW}, 
and in particular to the understanding of geometric scaling.
But no analytical solution could be found.

Important progress was accomplished when the first terms of the
large rapidity asymptotic expansion of 
the solutions to the Balitsky-Kovchegov equation were computed by
Mueller and Triantafyllopoulos \cite{MT}.
This expansion was subsequently systematized with 
the discovery \cite{MP2003,MP2004} 
that the BK equation lies in the universality class of the  
Fisher-Kolmogorov-Petrovsky-Piscounov (FKPP) equation \cite{FKPP,VS},
and that geometric scaling was equivalent to the property that the latter
admits traveling wave solutions.
The FKPP equation describes a vast class of physical phenomena, 
actively studied by
mathematicians and statistical physicists.

Meanwhile, Mueller and Shoshi \cite{MS} made a first step
in going beyond the BK approximation
to high energy scattering in QCD. Through a sophisticated analysis of the
unitarity and symmetry constraints on the QCD evolution, they were able 
to propose a correction to the
saturation scale, induced by effects 
neglected in the BK evolution,
and realized that geometric scaling was asymptotically broken.

The complete equivalence between
high energy QCD and statistical physics models of reaction-diffusion type, 
first understood by one of us (S.M.) \cite{M2005,IMM}, 
provided a new picture in which the nature of the
effects neglected by the BK equation
became transparent. The universality class of high energy QCD was identified as that
of the stochastic  Fisher-Kolmogorov-Petrovsky-Piscounov 
(sFKPP) equation \cite{Panja}.
The latter applies to a very wide class of physical phenomena, ranging
from directed percolation to population growth. It has many applications,
e.g. to biology, genetics, anthropology, fluid mechanics.
Roughly speaking, most of the physical situations 
in which objects evolve by multiplying
and diffusing, up to a
certain limiting threshold, may be captured by 
an equation in the universality class of the sFKPP equation.

The crucial point is that the 
QCD parton model has such dynamics, when viewed in a particular way that
we will explain in this paper.
Once this observation is made, all mathematical results
can be transposed to the physical context of high energy QCD \cite{IMM}.
In particular, Mueller and Shoshi's results on the saturation scale are recovered.
The breaking of geometric scaling, that they had understood qualitatively
in their approach, is shown to be related to the
statistical dispersion in the occupation numbers of partons in the 
hadronic Fock state.
This new point of view on high energy QCD not only confirmed and
brought new exact results
on QCD scattering amplitudes: it also 
provided a physical understanding of the very nature and the role of
fluctuations contained in the QCD evolution equations. Moreover, it
helped to understand how to set up a systematics to go beyond the BK equation.
It was recognized that the B-JIMWLK formalism was not 
complete~\cite{IT,MSW}, thus confirming what was
expected \cite{K,KMW,KL}. The form of
the new terms can be written down taking advantage of the
correspondence to statistical physics.

The goal of this paper is to provide numerical solutions of
the evolution equations of QCD, in order to study accurately 
their universal properties.
The outline goes as follows: in Sec.~\ref{sec_setup}, we 
explain how the parton model is related to a reaction-diffusion process.
>From physical but rigorous arguments,
we derive in a simple way the evolution
of the parton densities with energy, reviewing and expanding upon 
Refs.\ \cite{M2005,IMM}. In particular, we show how the
sFKPP equation appears.
In Sec.~\ref{sec_meanfield}, we study numerically in detail a useful
mean field approximation to the full evolution equations---that is, the BK equation---in relation
to the most recent analytical results.
In the final Sec.~\ref{sec_stochastic}, we go beyond, investigating
in a toy model
the full content of the sFKPP equation and its consequences in QCD.


\section{\label{sec_setup}Unitary evolution of amplitudes in the parton model}

In this first section, we recall the physical picture of
high energy scattering in the parton model (Sec.~\ref{subsec_pm}) 
insisting on the manifest link
to reaction-diffusion models. In a second step (Sec.~\ref{subsec_dm}), 
following Ref.~\cite{IMM}, 
we derive the general evolution equations in QCD.

\subsection{\label{subsec_pm}High energy scattering as a reaction-diffusion process}

\begin{figure}[th]
\begin{center}
\epsfig{file=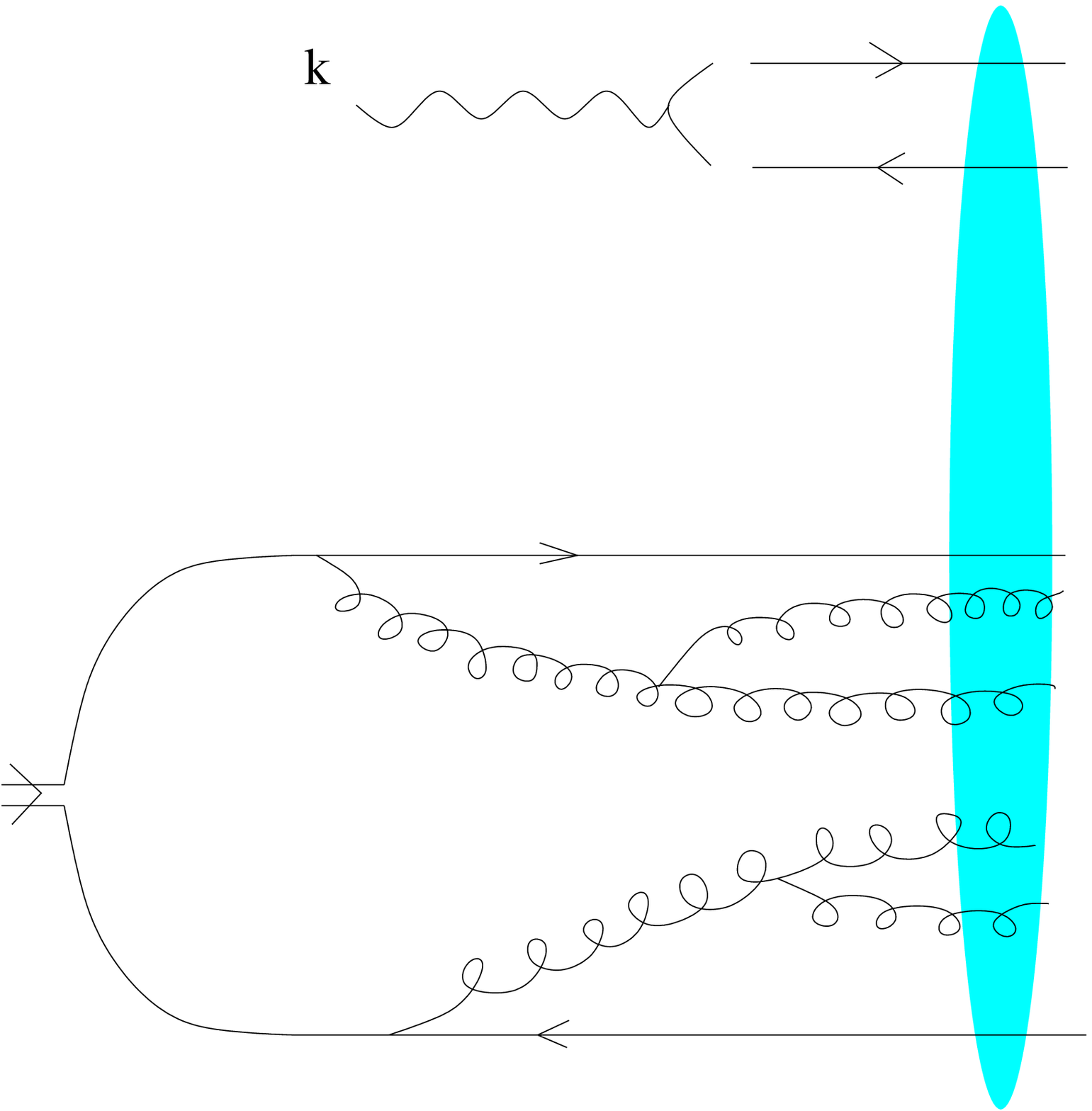,width=8cm}
\caption{\label{fluctuations}
The scattering of a probe (here, a virtual photon characterized
by the momentum scale $k$ interacting
through its $q\bar q$ Fock state) off a particular
4-gluon fluctuation of the target dipole.
}
\end{center}
\end{figure}

Let us consider the scattering of a hadronic probe 
off a given target, in the rest frame of the probe and at a 
fixed impact parameter.
In the parton model, the target interacts through one of its quantum
fluctuations (see Fig.~\ref{fluctuations}). 
The probe effectively ``counts'' the partons
in the current Fock state of the target
whose transverse momenta $k$ match the one of the probe:
the amplitude $T(k)$
for the scattering off this particular partonic
configuration is proportional to the number of partons $n(k)$.

In QCD, the wave function of a hadronic object 
is built up from successive splittings
of partons starting from the valence
structure.
As one increases the rapidity $Y$ by boosting the target,
the opening of the phase space for parton splittings
makes the probability for high occupation numbers larger.
In the initial stages of the evolution, the parton
density grows diffusively from these splittings.

On the other hand, the number of partons in each cell of transverse 
phase space is effectively limited 
to a maximum number $N$ that depends on 
the strength of the interaction of the probe with the partons,
that is on the relationship between $T$ and $n$. This property is necessary from
unitarity, which imposes an upper bound on the amplitude (e.g. $T(k)\leq 1$ with
standard normalizations) which, in turn, results in an
upper bound on $n(k)$. 
This is parton saturation and
is realized by a recombination process.

Viewed in this way, scattering in the 
parton model is a reaction-diffusion process. 
The rapidity evolution of the Fock states of the target hadron
is like the time evolution of a set of particles that diffuse in space, 
multiply and recombine
so that there is no more than $N$ particles
on each site, on the average. We parametrize space by 
the variable $x$: it will be related to $k$ in QCD.
The QCD amplitude $T$ is like the fractional occupation number
$u=n/N$ in the reaction-diffusion model. 

In the continuum limit,
it is known on general grounds \cite{DP} 
that, up to a change of variable, $u$ 
obeys the evolution equation
\begin{equation}
\partial_t u(x,t)=D\partial_x^2 u(x,t)+\lambda u(x,t)-\mu u^2(x,t)
+\frac{1}{\sqrt{N}}\sqrt{2u(x,t)}\,\eta(x,t)\ ,
\label{RFT}
\end{equation}
where $\eta$ is a Gaussian white noise, that is, a random function satisfying
\begin{equation}
\langle\eta(x,t)\rangle=0\ \ \ \mbox{and}\ \ \
\langle\eta(x,t)\eta(y,t^\prime)\rangle=\delta(t-t^\prime)\delta(x-y)\ .
\label{noise}
\end{equation}
Eq.~(\ref{RFT}) is known as the ``Reggeon field theory'' equation.
It is to be interpreted in the standard It\={o} way, as the continuum limit
of a discretized equation.
It belongs
to the same universality class as
the stochastic Fisher-Kolmogorov-Petrovsky-Piscounov (sFKPP)
equation (which may be obtained by
the replacement $\sqrt{u}\rightarrow\sqrt{u(1-u)}$ in Eq.~(\ref{RFT})).
$D$ and $\lambda$ are parameters that characterize 
the diffusive growth of the particles.
The structure of Eq.~(\ref{RFT}) is very transparent: the first two terms
represent the growing diffusion of the particles, characterized by the parameters
$D$ and $\lambda$, the third term implements the recombination process and
tames this growth when maximum
occupancy is approached. The last term results from the
stochastic dynamics, whose effect reflects the finiteness of the number of
particles.
Note that this term gives rise to fluctuations in $u(x,t)$ 
and, consequently, in the particle
number $n$,  that are proportional to $\delta n\propto\sqrt{n}$.
This is typical of fluctuations of independent random numbers, 
and reflects well the statistical origin of this term.

We do not mean that Eq.~(\ref{RFT}) fully represents parton evolution.
Indeed, in the parton model, we anticipate that the rapidity evolution
of parton densities be given, in the linear regime, by the BFKL equation,
which is more complicated than a second order differential operator.
But what we mean is that Eq.~(\ref{RFT}) describes exactly the critical
behavior of the QCD amplitude $T\ (\sim u)$, that is its 
large rapidity (time) and large maximum
occupancy number $N$ limit, up to the replacement of
the relevant parameters.

All higher order operators beyond the ones appearing in 
Eq.~(\ref{RFT}) are irrelevant in these limits.
As for the linear part, this is clear: the large time behavior stems from
a saddle point that selects the quadratic approximation
to the growing diffusion kernel. As a matter of fact, any equation of the form
\begin{equation}
\partial_t u(x,t)=K\otimes u(x,t)-\mu u^2(x,t)
+\frac{1}{\sqrt{N}}\sqrt{2u(x,t)}\,\eta(x,t)\ ,
\label{RFTK}
\end{equation}
where $K$ is an appropriate differential or integral kernel 
encoding the growing diffusion,
is believed to belong to
that universality class and hence to exhibit the same asymptotic solutions.
By ``appropriate'' we mean in practice
that the phase velocity 
\begin{equation}
v(\gamma)=\frac{K(\gamma)}{\gamma},\ \ \mbox{with}
\ \ K(\gamma)=e^{\gamma x} (K\otimes e^{-\gamma x}) 
\label{vgamma}
\end{equation}
of a wave of wave number $\gamma$ propagated by the linear part
of Eq.~(\ref{RFTK}),
must be a function of $\gamma$ that has a minimum in its domain of
analyticity, but we should also stress that
a general mathematical definition of the universality class
of the Reggeon field equation is still not 
available~\cite{Panja}.\footnote{Strictly speaking, 
the case of the running coupling
is not captured by Eq.~(\ref{RFTK}) since $K$ would
not be a proper diffusion kernel. However, it is possible to treat
the running coupling through a
generalization of the methods known to treat the sFKPP 
equation, as was done
in the mean field case in Ref.~\cite{MP2003}.}
As for the nonlinear part in Eqs.~(\ref{RFT}) and (\ref{RFTK}), 
terms of the form $u^j$ for
$j>2$ would have no influence because the region $u\sim 1$ in which
these terms show up is absorptive \cite{VS}. 
Furthermore, we will see that the time evolution
is essentially driven by the region where $u$ is small. 
Thanks to this property, we will not have to
bother about the exact way how parton saturation actually occurs in QCD:
it is enough to know that there is an upper bound on the number of partons
by unit of transverse phase space.

So far, we have focused on the evolution of one (partonic) realization,
described by Eq.~(\ref{RFTK}). In the QCD context, $u\sim T$ 
represents the scattering amplitude off one particular Fock state.
It is not a physical observable because 
it is not possible to select a particular Fock state of the target
 in a particle physics experiment.
The physical amplitude that is measured in an experiment 
is the average over all accessible 
random (partonic) realizations. Only a few moments 
of $T$ may be measurable:
its average, related to total cross sections, and its variance, related
to diffractive cross sections, namely:
\begin{equation}
\begin{split}
\sigma_{tot}&\sim \langle T \rangle\\
\sigma_{diff}&\sim \langle T^2 \rangle-\langle T \rangle^2\ .
\end{split}
\label{crosssections}
\end{equation}
The evolution of the moments of $T$ (or of $u$) may be computed 
directly from Eq.~(\ref{RFT}) (or Eq.~(\ref{RFTK})). 
Taking the average of Eq.~(\ref{RFTK}), with the help of Eq.~(\ref{noise}),
it is straightforward to obtain the evolution of the latter
\begin{subequations}\label{hierarchy}
\begin{equation}
\partial_t \langle u(x,t)\rangle =K\otimes \langle u(x,t)\rangle
-\mu \langle u^2(x,t)\rangle\ .
\label{hierarchy1}
\end{equation}
However, this equation is not closed: it calls for an 
evolution equation for the correlator
$\langle u^2(x,t)\rangle$, that can be derived again 
from Eq.~(\ref{RFTK}) with the help of Eq.~(\ref{noise}):
\begin{equation}
\partial_t \langle u(x,t)u(y,t)\rangle =(K_x+K_y)\otimes\langle u(x,t)u(y,t)\rangle
-\mu \langle u^2(x,t)u(y,t)\rangle-\mu \langle u(x,t)u^2(y,t)\rangle
+\frac{2}{N}\delta(x\!-\!y)\langle u(x,t)\rangle\ ,
\label{hierarchy2}
\end{equation}
\end{subequations}
and so on for the higher order correlators, leading eventually to an
infinite hierarchy of coupled equations.

Note that the hierarchical form~(\ref{hierarchy})
was proposed by Parisi and Zhang \cite{Parisi} to describe 
a growth process known as
the Eden model. In the same paper, they also obtained 
the same hierarchy from the Reggeon action, 
which had already been proved to be related to
stochastic processes \cite{ACMP_GS}.
Only the boundary terms (last term in the right-hand side of Eq.~(\ref{hierarchy2}))
could not be obtained
from the Reggeon action through the Parisi-Zhang procedure: 
indeed, they refer explicitly to the way one counts the
objects, as can be seen from the fact that this term is
proportional to $2/N$, and the knowledge of the evolution encoded in the
bulk terms of the action they had used is not enough.

Due to universality of the asymptotics of reaction-diffusion processes,
it is straightforward to take over Eq.~(\ref{RFTK}) or
equivalently Eqs.~(\ref{hierarchy1}) and (\ref{hierarchy2}) 
(and the higher equations in this hierarchy),
together with their {\it exact} asymptotic solutions that will be exhibited
in the following sections, to describe high energy QCD.
Basically, an elementary analysis of the parton splitting process and
how the partons interact perturbatively with a probe will be enough.
This will enable us to identify the variables $t$ and $x$, the kernel
of the linear evolution $K$ and the strength of the noise $1/N$
that correspond to the QCD parton model.
We will turn to this analysis in the next section.

For Eq.~(\ref{RFTK}), being both explicitly nonlinear and stochastic, it proves
a formidable task to find solutions.
However, there has been much progress recently 
in this direction~\cite{Panja}.
It was realized that, starting from a localized initial condition,
the large time solution is a traveling wave moving in the direction of larger $x$,
up to some noise.
The position $X_t$ of that wave 
may be defined in several ways. For example, one can pick
a constant $\kappa$, and define $X_t$ in such a way that
\begin{equation}
u(X_t,t)=\kappa\ .
\label{defpos1}
\end{equation}
An alternative definition, useful in particular in the case of discrete models,
could be
\begin{equation}
X_t=\int_0^\infty dx\,u(x,t)\ .
\label{defpos2}
\end{equation}
In the context of QCD, $X_t$ is related to the saturation scale, that 
is, the point at which parton saturation effects set in.

Outstanding results have been obtained on $X_t$ and on the profile of
the wave front $u(x,t)$ in the region $1/N\ll u\ll 1$.
An interesting point is that the analytical formulas depend only
on some local properties of
the characteristic function $K(\gamma)$ 
of the linear diffusion kernel.


\subsection{\label{subsec_dm}The evolution equations of the QCD parton model}

In this section, we review the discussion of Ref.~\cite{IMM}.
Our goal is to identify the relevant variables $t$, $x$, $N$
and the evolution kernel $K$ that correspond to the physical situation
of QCD.

We consider the scattering of a dipole of variable size $r$ (the
probe) off a dipole of size $r_0$ (the target). A natural variable
that will be used throughout is $\rho\!=\!\ln(r_0^2/r^2)$. We go
to the rest frame of the probe so that the target carries all the
available rapidity $Y$. The impact parameter $b$ between the
dipoles is fixed.

At high energy, the quantum fluctuations of the target
that interact with the probe
are dominated by gluons.
It proves useful to represent this set of
partons by color dipoles \cite{CD}.
This is possible in the large-$N_c$ limit, where
gluons are similar to zero-size $q\bar q$ pairs
and non-planar diagrams are suppressed.
Subleading-$N_c$ terms do not contribute 
to the evolution of the scattering
amplitudes when they are small. However,
the dipole approximation breaks down when the amplitudes
approach their unitarity limits at the same time as nonlinearities
set in, but as we will see,
that does not hamper getting the
right asymptotics for the physical
quantities that we are going to compute.

We denote by $T(r,Y)$ the scattering amplitude of the probe off
a given partonic realization $|\omega\rangle$
of the target, obtained after a rapidity evolution $Y$
(the dependence on $b$ is understood).
It is a random variable, whose probability
distribution is related to the distribution of the different
Fock state realizations
of the target. The values of $T(r,Y)$ range between 0 (weak interaction)
and 1 (unitarity limit).
$T(r,Y)$ will be an essential intermediate quantity in our
calculations, but as has been explained in Sec.~\ref{subsec_pm},
it is not an observable.
The physical dipole-dipole scattering amplitude $A(r,Y)$
is the statistical average over
all partonic fluctuations of the target at rapidity $Y$, i.e.
\begin{equation}
A(r,Y)=\langle T(r,Y) \rangle ;
\end{equation}
see Eq.~(\ref{crosssections}).

When $T$ is small,  the scattering amplitude off one particular Fock state
$\omega(Y)$ is the sum of all elementary amplitudes with each of the dipoles,
i.e.
\begin{equation}
T(r,Y)=\sum_{i\in \omega(Y)} T_{el}(r,r_i)\ ,
\end{equation}
where $i$ labels the dipoles
in the Fock state of the target at the time of the interaction.
$T_{el}$ is the elementary dipole interaction and
is essentially local in impact parameter.
$T_{el}$ behaves like
\begin{equation}
T_{el}(r,r_i)\sim
\alpha_s^2 \,\frac{r_<^2}{r_>
^2}\ ,\ \mbox{with}\
r_<=\min(|r|,|r_i|)\ ,\ r_>=\max(|r|,|r_i|)
\label{Tel}
\end{equation}
when the dipoles overlap, and vanishes otherwise. We have
neglected ${\mathcal{O}}(1)$ factors and logarithms, but these
approximations 
do not affect the
results that we shall obtain, which are largely independent of the
details. Eq.~(\ref{Tel}) shows that
the amplitude $T(r,Y)$ is simply counting the number $n(r,Y)$
of dipoles of size $r$ within a disk of radius $r$ centered at the
impact parameter of the external dipole:
\begin{equation}\label{ndef}
T(r,Y)\sim \alpha_s^2\, n(r,Y)\ .
\end{equation}
Note that $n(r,Y)$ can take only discrete values:\ Indeed, it
represents the number of quanta of a particular realization of
the field of the target.
The unitarity bound on $T$ implies that $n(r,Y)$ is also constrained by an
upper bound $N\equiv 1/\alpha_s^2$.
Later on, we shall switch to momentum space by the Fourier transformation
$r\leftrightarrow k$. Of course, the relationship~(\ref{ndef})
will also hold for the Fourier conjugates.

We emphasize that the relationship~(\ref{ndef})
is only valid within the framework of the above-mentioned approximations.
A more accurate treatment of how the counting of the partons is actually
realized in QCD would eventually lead to more complex equations, 
see Ref.~\cite{MSW}. We will not enter these complications in this paper
since we are only interested in the study of the asymptotics that are common
with models from statistical physics, for which the counting rule~(\ref{ndef})
is precise enough.

The rapidity
evolution law for $\langle T(r,Y)\rangle$ is known
and we refer the reader to earlier literature for its derivation
(e.g.~\cite{IMM}).
It reads\footnote{The impact parameter dependence could
be easily put back in Eq.~(\ref{balitsky}).
We have omitted it for simplicity and since it is enough
for our purpose
to assume locality of the evolution.}
\begin{subequations}\label{balitsky}
\begin{equation}
\partial_Y \langle T(r,Y)\rangle
=\frac{\bar\alpha}{2\pi}\int d^2 z \frac{r^2}{z^2(r\!-\!z)^2}
(\langle T(z,Y)\rangle+\langle T(r\!-\!z,Y)\rangle
-\langle T(r,Y)\rangle
-\langle T(z,Y)T(r\!-\!z,Y)\rangle)\ .
\label{balitsky1}
\end{equation}
Eq.~(\ref{balitsky1}) is not a closed equation for $\langle T
\rangle$:\ It depends upon the correlator $\langle
T(z,Y)T(r\!-\!z,Y)\rangle$. 
An evolution equation for $\langle T(r_1,Y)T(r_2,Y)\rangle$
may be derived in the same way. One gets
\begin{multline}
\partial_Y \langle T(r_1,Y)T(r_2,Y)\rangle
=\frac{\bar\alpha}{2\pi}\int d^2 z \frac{r_1^2}{z^2(r_1\!-\!z)^2}
(
\langle T(z,Y)T(r_2,Y)\rangle+\langle T(r_1\!-\!z,Y)T(r_2,Y)\rangle
-\langle T(r_1,Y)T(r_2,Y)\rangle\\
-\langle T(z,Y)T(r_1\!-\!z,Y)T(r_2,Y) \rangle )\\
+\frac{\bar\alpha}{2\pi}\int d^2 z \frac{r_2^2}{z^2(r_2\!-\!z)^2}
(\langle T(r_1,Y)T(z,Y)\rangle+\langle T(r_1,Y)T(r_2\!-\!z,Y)\rangle
-\langle T(r_1,Y)T(r_2,Y)\rangle\\
-\langle T(r_1,Y)T(z,Y)T(r_2\!-\!z,Y) \rangle )\ ,
\label{balitsky2}
\end{multline}
\end{subequations}
which again calls for equations for higher correlators.
Eqs.~(\ref{balitsky})
turn out to be the first two parts of
an infinite hierarchy originally derived by Balitsky~\cite{B}, restricted
to dipoles.

In order to make the correspondence with reaction-diffusion models
more concrete at the physical level, it is useful at this point to
discuss qualitatively the typical shape of
$T(r,Y)$ from the evolution. 
(The reader may find the complete derivation in Ref.~\cite{IMM}.)
As one takes a new step in
rapidity, each of the dipoles $r_i$ already present in the wave
function from the previous step and for which $T(r_i,Y)\ll 1$
may split into two new dipoles. 
The most probable splittings are those 
in which both child dipoles have a size comparable to the size of
the parent dipole: they occur with probability one 
in each interval $\Delta Y\sim 1/\bar\alpha$. 
Thus the main mechanism for the rise of $T(r_i,Y)$ with $Y$
is a growing diffusion around the size of the initial dipole
$r_i$, that is encoded in the linear part of Eqs.~(\ref{balitsky}). 
The nonlinear terms appearing therein
tame this growth in order to satisfy the unitarity limits
on $\langle T\rangle$. 

In a typical partonic configuration as obtained
after a sufficiently large rapidity evolution, the
dipoles appear to be densely distributed around the size $r_0$ ($T(r,Y)$ is large),
but they become more rare with decreasing $r$ (or increasing $\rho$),
and for sufficiently large $\rho$ one meets only rare fluctuations
which involve
one (or few) dipoles and for which $T(r)\simeq\alpha_s^2$.
It is a wave front which with increasing $Y$
progresses towards larger values of $\rho$. 
The position of that front is characterized by a momentum scale
$Q_s(Y)$ called the saturation momentum.
It is natural to define it e.g. by the value of the inverse dipole
size for which the amplitude reaches some predefined
number $\kappa$, i.e.
$T(r\!=\!1/Q_s(Y),Y)=\kappa$: this corresponds to 
prescription~(\ref{defpos1}).

We now proceed to the identification 
of the QCD evolution equation
to the Reggeon field theory
equation at the technical level. 
Following Ref.~\cite{K}, we perform the transformation
\begin{equation}
T(k,Y)=\int_0^{\infty}\frac{dr}{r}J_0(k r)T(r,Y)
\end{equation}
to momentum space.
The main positive outcome of such a transformation is
to make the nonlinear terms
in Eqs.~(\ref{balitsky}) local in $k$, while obviously the linear
part stays an integral kernel.
It leads to the formal simplification
\begin{equation}
\left\{
\begin{aligned}
\partial_{\bar\alpha Y} \langle T(k,Y)\rangle
&=\chi(-\partial_L) \langle T(k,Y)\rangle
-
\langle T^2(k,Y)\rangle \\
\partial_{\bar\alpha Y} \langle T(k_1,Y)T(k_2,Y)\rangle
&=(\chi(-\partial_{L_1})+\chi(-\partial_{L_2})) 
\langle T(k_1,Y)T(k_2,Y)\rangle\\
&\phantom{TTT}-\langle T^2(k_1,Y)T(k_2,Y)\rangle
-\langle T(k_1,Y)T^2(k_2,Y)\rangle\\
 &\cdots
\end{aligned}
\right.
\label{balitskyk}
\end{equation}
where $L\equiv \ln (k^2/\Lambda^2)$ and
$\chi(\gamma)=2\psi(1)-\psi(\gamma)-\psi(1\!-\!\gamma)$ is
the eigenvalue of the BFKL kernel.
By $\chi\left(-\partial_L\right)$, we mean the BFKL integro-differential
operator that may be defined 
with the help of the formal series expansion
\begin{equation}
\chi\left(-\partial_L\right)=
\chi(\gamma_0){\mathbbm 1}+\chi ^\prime(\gamma_0)(-\partial_L-\gamma_0{\mathbbm 1})
+{\scriptstyle\frac12}\chi ^{\prime\prime}(\gamma_0)
(-\partial_L-\gamma_0{\mathbbm 1})^2
+
\cdots
\label{chiexpansion}
\end{equation}
for some given $\gamma_0$ between 0 and 1, i.e. for the principal branch
of the function $\chi$.

It is clear that, as it stands,
the hierarchy (\ref{balitskyk}) admits the
factorized ``mean field'' solution \cite{trivsol}
\begin{equation}
\langle T(k_1,Y)\cdots T(k_n,Y)\rangle=
\lambda^{n-1} \langle T(k_1,Y)\rangle\cdots \langle T(k_n,Y)\rangle
\label{trivialsol}
\end{equation}
where $\lambda$ is any constant.
However, this is not the physical solution that corresponds to
scattering in QCD.
The set of equations~(\ref{balitskyk}) may be supplemented by
boundary terms if one wants to represent explicitly the
elementary dipole interaction (i.e. the ``counting rule''), 
which is absent in 
Eq.~(\ref{balitskyk}) (compare to Eq.~(\ref{hierarchy2})).
By analogy with statistical mechanics (see Eq.~(\ref{hierarchy2})),
this term should be \cite{Parisi}
\begin{equation}
2\alpha_s^2\delta(L_1-L_2)\langle T(k_1,Y)\rangle
\label{sourceterm}
\end{equation}
in the second equation.
With this term, it is clear that~(\ref{trivialsol}) does no
longer solve the hierarchy.
Our method of derivation does not allow to find directly
this boundary term, exactly for the same reasons Parisi and Zhang did not get it in their derivation of the
hierarchy from the Reggeon action \cite{Parisi}.
Eq.~(\ref{sourceterm}) is nonzero when the transverse momenta of the two dipoles
in the Fock space of the probe are equal, and thus, may scatter off the same
dipole in the Fock space of the target:\ This was neglected in our derivation
and has to be reintroduced by hand.

The evolution equation for $T$ may also be written
in the form of a stochastic differential equation, namely,
\begin{equation}
\partial_{\bar\alpha Y}T(k,Y)=\chi(-\partial_L) T(k,Y)-T^2(k,Y)+
\alpha_s\sqrt{2T(k,Y)}\eta(k,Y)
\label{stochastic}
\end{equation}
that gives back the hierarchy~(\ref{balitskyk}) 
for the correlators,
supplemented with 
boundary terms such as~(\ref{sourceterm}).
Such a form was proposed for the first time in Refs.~\cite{M2005}
on physical grounds:\ Indeed, this is the universal asymptotic form of the evolution
of any reaction-diffusion type process \cite{DP}.
Subsequently, it was shown on the technical level
how the B-JIMWLK formalism can be made consistent with these physics~\cite{IT,MSW}, 
and how this term
may describe Pomeron loops
(see also Refs.~\cite{Lpomeronloop}). The source term~(\ref{sourceterm})
corresponds exactly to the one found there.

The identification with the Reggeon field theory equation is now straightforward:
\begin{equation}
\begin{split}
x&\rightarrow L
\ ,\ \
t\rightarrow \bar\alpha Y\ ,\\
K&\rightarrow\chi(-\partial_L)\ ,\ \ \mu\rightarrow 1\ ,\\
N&\rightarrow 1/\alpha_s^2\ .
\end{split}
\end{equation}
Let us discuss the validity of Eq.~(\ref{stochastic}) 
when applied to high energy scattering,
and the physical
expectations.
As stated before, it is not expected to be exact, but 
we believe that it captures the
essential physics of the QCD parton model.
Since the linear part drives the front propagation at 
high rapidity, and since it is taken exactly into account here
(to leading order in $\bar\alpha_s Y$), 
we expect that the solutions to this
equation match the asymptotics of full QCD as far
as the position of the front is concerned. This means 
that one should be able to get the asymptotics
of the moments of the saturation scale $Q_s$
for $Y$ large, $\alpha_s\rightarrow 0$, 
and at a fixed impact parameter
by solving~(\ref{stochastic}).
The latter limitations are necessary for the counting
rule~(\ref{ndef}) to be valid.
Similarly, the shape of $T$  in the region $1/N\ll T\ll 1$
is known to be universal and hence should also be obtainable from
Eq.~(\ref{stochastic}).

Of course, strictly speaking, these statements
have the status of a conjecture,
that will have to be confirmed by accurate numerical calculations,
and in the long term, by a better mathematical understanding 
of the solutions
of equations of the form~(\ref{stochastic}). 

There is still some important
difference between the original Reggeon field
theory equation~(\ref{RFT}) 
and the equation that we have obtained for high energy
QCD. Indeed,
$T(k,Y)=1$ is not a fixed point of the QCD evolution,
unlike the case of the original Reggeon field theory equation, so the
solutions cannot match in that region.
Actually, it is known that $T(k,Y)\sim \ln(Q_s^2/k^2)$ in the region
$T\ge 1$ \cite{KL2005}.
But anyway, we would not solve Eq.~(\ref{stochastic}) around $T\ge 1$
because in the QCD case, the nonlinearity cannot be reduced to a
simple quadratic term as in ~(\ref{stochastic}), since 
also color structures beyond dipoles are expected to play a role
there \cite{KL}. 
That does presumably not influence the asymptotic saturation
scale, which is completely determined by the linear part of the
evolution equation,
but it certainly modifies the shape of $T$ and of its correlators
close to the unitarity limits.


\section{\label{sec_meanfield}The mean field picture}

The mean field approximation
$\langle T^2(k,Y)\rangle\simeq \langle T(k,Y)\rangle^2\equiv A^2(k,Y)$  casts Eq.~(\ref{balitsky}) into a
closed form, known as the Balitsky-Kovchegov (BK) 
equation \cite{B,K}:
\begin{equation}
{\partial_{\bar\alpha Y}}{A(k,Y)}=
\chi\left(-\partial_L\right){A(k,Y)}
-{A^2(k,Y)}\ .
\label{bk}
\end{equation}
A priori, the validity condition for
such an approximation is not clear. We will be able to
assess it only by studying the full problem including fluctuations.
It is nevertheless useful to start with such an approximation
because it is tractable both analytically and numerically,
and because the full stochastic solution
is nothing else than a perturbation of the mean field solution.


\subsection{\label{sec:general}General properties of the BK equation}

It has recently been understood \cite{MP2003} that the BK equation belongs to the
universality class of the FKPP equation, and as such,
it admits a family of 
traveling wave solutions \cite{Bramson}. That is, there exists a function
of the rapidity $Q_s(Y)$ such that
\begin{equation}
{ A}(k,Y)={ A}(L-\ln Q_s^2(Y)) 
\label{twave}
\end{equation}
is a solution to Eq.~(\ref{bk}). 
For this family of solutions,
$A(\xi)$ on the r.h.s. is essentially a slowly varying
function of $\xi\equiv L-\ln Q_s^2(Y)$ (tending to a 
constant in the case of the pure FKPP equation)
for $\xi<0$, and is exponentially decreasing at large $\xi$
\begin{equation}
{A}(\xi)\sim e^{-\gamma \xi}\ .
\label{shape}
\end{equation}
$Q_s$ is the saturation scale introduced before, that is, the
transition point between the linear and the saturation regime.
Note that the general traveling wave property~(\ref{twave}) is
equivalent to geometric scaling, a feature of the
data for $\gamma^* p$ scattering at high energy discovered a few years ago
\cite{SGK}.

The crucial point is that the properties of these traveling waves
are to a large extent determined by the large-$k$ tail,
where the amplitude ${ A}$ is small and thus where
Eq.~(\ref{bk}) may be linearized. This very important property
is due to the particular
propagation mode of the front, which is said to be ``pulled'' along by
its tail.

The linear part of the BK equation~(\ref{bk}) 
has the characteristic function (see Eq.~(\ref{vgamma}))
\begin{equation}
v(\gamma)=\frac{\chi(\gamma)}{\gamma}\ .
\label{dispersion}
\end{equation}
$v(\gamma)$ is the phase velocity of a wave 
of wave number $\gamma$.
It has a minimum at $\gamma_0 \approx 0.627$.

Starting from a given initial condition ${ A}_0$ which behaves asymptotically
like $e^{-\beta\xi}$, there are two relevant cases.
Either $\beta<\gamma_0$, in which case the large-$Y$ 
asymptotic solution conserves $\gamma=\beta$, or $\beta\geq\gamma_0$, in which
case the wave front will converge asymptotically to the shape $e^{-\gamma_0\xi}$,
that moves at velocity $d\ln Q_s^2/dY=\bar\alpha v(\gamma_0)$.
This property can be understood in a simple way.
The wave packet that corresponds to a physical initial condition,
which has at most a mild growth as $\xi\rightarrow -\infty$  
and that decreases like
$e^{-\beta\xi}$ for  $\xi\rightarrow +\infty$,
contains all waves of wave number $\gamma$ ranging from $-\infty$ to $\beta$.
At large times, the slowest of these waves will determine the velocity of
the wave front. The slowest wave is either $\beta$ itself if $\beta<\gamma_0$, or
$\gamma_0$ in the opposite case.

The latter case is the physically relevant one for the
BK equation \cite{MP2004}:\ Indeed, color transparency 
implies that for large momenta, i.e.
large values of $\xi$, the QCD amplitudes behave like $e^{-\xi}$.
Thus $\beta=1$, which is larger than $\gamma_0$.

The transition from the initial condition at $Y\!=\!0$ to the
asymptotic traveling wave induces corrections to the velocity of
the front (i.e. to $Q_s$) and to its shape.
The first few orders in a $Y$-expansion of $\ln Q_s^2$
are completely determined by 
Eq.~(\ref{dispersion}):
\begin{equation}
\frac{d}{dY}\ln Q_s^2(Y)
=\bar\alpha\frac{\chi(\gamma_0)}{\gamma_0}-\frac{3}{2\gamma_0}\frac{1}{Y}
+\frac{3}{2\gamma_0^2}
\sqrt{\frac{2\pi}{\bar\alpha\chi^{\prime\prime}(\gamma_0)}}\frac{1}{{Y}^{3/2}}
+{ O}(1/Y)\ .
\label{satscalnl}
\end{equation}
The first term on the right-hand side was first computed in Ref.~\cite{GLR} in the
context of the GLR equation. The second term was found in Ref.~\cite{MT}, while the
last one was derived for QCD in Ref.~\cite{MP2004}.

The front itself has a rapidity expansion
around its asymptotic shape~(\ref{shape}) which reads, to the same
order of approximation
\begin{equation}
A(k,Y)=\left(\frac{k^2}{Q_s^2(Y)}\right)^{-\gamma_0}\times \psi(k,Y)\ ,
\label{defredfront}
\end{equation}
where the reduced front $\psi(k,Y)$ was computed in Ref.~\cite{MP2004}:
\begin{multline}
\psi(k,Y)=C_1
e^{-z^2}
\times\Bigg\{\gamma_0\ln [k^2/{Q_s^2(Y)}]+C_2
+\left(3-2C_2+\frac{\gamma_0\chi^{(3)}(\gamma_0)}{\chi^{\prime\prime}(\gamma_0)}\right)
z^2\\
-\Bigg({\frac23}\frac{\gamma_0\chi^{(3)}(\gamma_0)}{\chi^{\prime\prime}(\gamma_0)}
+
\frac13{}_2\!F_2\left[1,1;{\scriptstyle\frac52},3;
z^2\right]\Bigg)z^4
+6\sqrt{\pi}
\left(1-{}_1\!F_1\left[-{\scriptstyle \frac12},{\scriptstyle \frac32};
z^2
\right]\right)z+{ O}(1/\sqrt{Y})
\Bigg\}\ .
\label{frontnl}
\end{multline}
$z\equiv\ln(k^2/Q_s^2(Y))/{\sqrt{2\bar\alpha \chi^{\prime\prime}(\gamma_0) Y}}$
is the so-called ``leading edge'' variable~\cite{VS}.

The dominant term (first factor in the right-hand side)
is the exponential~(\ref{shape}) corrected by a linear factor that
represents absorption, and by a Gaussian
whose width increases with rapidity, namely
\begin{equation}
{\psi}_{LO}(k,Y)=C_1
\left(\gc\ln\left(\frac{k^2}{Q_s^2(Y)}\right)+C_2\right) 
\exp\left(-\frac{\ln^2 ({k^2}/{Q_s^2(Y)})}{2\bar\alpha\xggc Y}
\right)\ .
\label{eq-redfront}
\end{equation}
Having an explicit $Y$ dependence, this factor clearly violates geometric
scaling.
One can estimate from Eq.~(\ref{eq-redfront}) 
how many steps in rapidity $\Delta Y$ are needed in order for
the exponential shape to set in 
within a ``distance'' $\ln k^2-\ln Q_s^2$
from the bulk of the front: it
corresponds to the point where $z$ approaches 1, i.e.
\begin{equation}
\bar\alpha \Delta Y\sim \frac{\ln^2 (k^2/Q_s^2)}{2\chi''(\gamma_0)}\ .
\label{diffusiontmf}
\end{equation}


\subsection{Numerical implementation and approximation schemes}

Numerical studies of the BK equation were pioneered in Refs.~\cite{AB,L2001,GMS,AAMSW}.
These works anticipated qualitatively the traveling wave behavior of the
large rapidity solution.
In this paper, we will focus on the quantitative comparison with the most
recent analytical results.

For the purpose of numerical implementation, the following form 
of the BFKL kernel in Eq.~(\ref{bk})
is well-suited
\begin{equation}
\chi(-\partial_L) T(L)=\int_{-\infty}^{+\infty}dL^\prime
\left(
\frac{T(L^\prime)-e^{L-L^\prime}T(L)}{\left|
1-e^{L-L^\prime}
\right|}
+
\frac{e^{L-L^\prime}T(L)}
{\sqrt{4+e^{2(L-L^\prime)}}}
\right)\ .
\label{numBKkernel}
\end{equation}
There are several known approximations to this kernel.

In the so-called diffusive approximation, one keeps only terms up to second order
in the expansion~(\ref{chiexpansion}), which makes the BFKL kernel
local:
\begin{equation}
\chi(\gamma)\simeq
\chi(\gamma_0)+\chi ^\prime(\gamma_0)(\gamma-\gamma_0)
+{\scriptstyle\frac12}\chi ^{\prime\prime}(\gamma_0)
(\gamma-\gamma_0)^2\ .
\label{chiexpansion2}
\end{equation}
Within this approximation, the BK equation becomes a partial differential equation, 
which is exactly equivalent to the FKPP equation up to a trivial change of variables
\cite{MP2003}.
Moreover, these terms are enough to yield the exact asymptotics
of Eqs.~(\ref{satscalnl},\ref{frontnl}) to that order.
For the leading order correction, it is clear
because the expansion~(\ref{chiexpansion2}) 
is equivalent to a saddle point approximation.
The next-to-leading order only depends on $\chi^{\prime\prime}(\gamma_0)$,
but this is probably an accident, since a priori one would
expect $\chi^{\prime\prime\prime}(\gamma_0)$ to also play a role.

One can also take a different point of view, and focus instead on the pole
structure of the BFKL kernel, by starting from its meromorphic expansion.
An accurate representation of the principal
branch of the $\chi$ function consists in keeping only
the two poles at $\gamma=0$ and $\gamma=1$:
\begin{equation}
\chi(\gamma)\simeq
\frac{1}{\gamma}+\frac{1}{1-\gamma}+4(\ln 2 -1)\ ,
\end{equation}
where the constant is adjusted in such a way that 
$\chi({\scriptstyle\frac12})$ coincides with its exact value. 
Within this scheme, 
$\chi(-\partial_L)$ is an integral operator that admits the representation
\begin{equation}
\chi(-\partial_L)T(L)\simeq\int_L^{+\infty}dL^\prime\, T(L^\prime)
+\int_{-\infty}^L dL^\prime\, e^{L^\prime-L}\, T(L^\prime)
+4(\ln 2-1) T(L)\ .
\label{num2pkernel}
\end{equation}

We have solved numerically 
the BK equation with these kernels using 
several different methods, which has allowed cross-checking of the accuracy and 
reliability. 
The main algorithm (also used in Ref.~\cite{Enberg:2005et}) is based on
Chebyshev approximation of the integrand.\footnote{The code can be
downloaded
from \texttt{http://www.isv.uu.se/\~{}enberg/BK/}} The integrand is defined
on a grid in momentum space given by the extrema of the Chebyshev
polynomials.
The integral is then computed by using a discrete cosine transform, and the
ensuing system of ordinary differential equations is solved using a
fourth-order Runge--Kutta method. This makes the program very efficient,
and
makes it possible to discretize the system on quite a small grid; for
most of
the results in this paper, a grid size of 256 points was sufficient,
with the
$L$ variable ranging between $L_\text{min}=-20$ and $L_\text{max}=138$.

We will now focus on the numerical results and their physical interpretation.


\subsection{Front formation and wave propagation}

As an initial condition for the evolution, 
following Ref.~\cite{AAKSW},
we assume the McLerran--Venugopalan 
(MV) model \cite{MV}, defined in coordinate space by
\begin{equation}
{T}(r,Y\!=\!0)=\exp\left(-\frac14{r^2 Q_s^2(Y\!=\!0)}
\ln\left[e+1/(r^2\Lambda^2)\right]\right)\ ,
\label{MV}
\end{equation}
with starting scale $Q_s^2(Y=0)$ 
set to $1\ \mbox{GeV}^2$ and $\Lambda=200\ \mbox{MeV}$.
The coupling constant $\bar\alpha$ is frozen at 0.2 for all our
calculations.

\begin{figure}[th]
\begin{center}
\epsfig{file=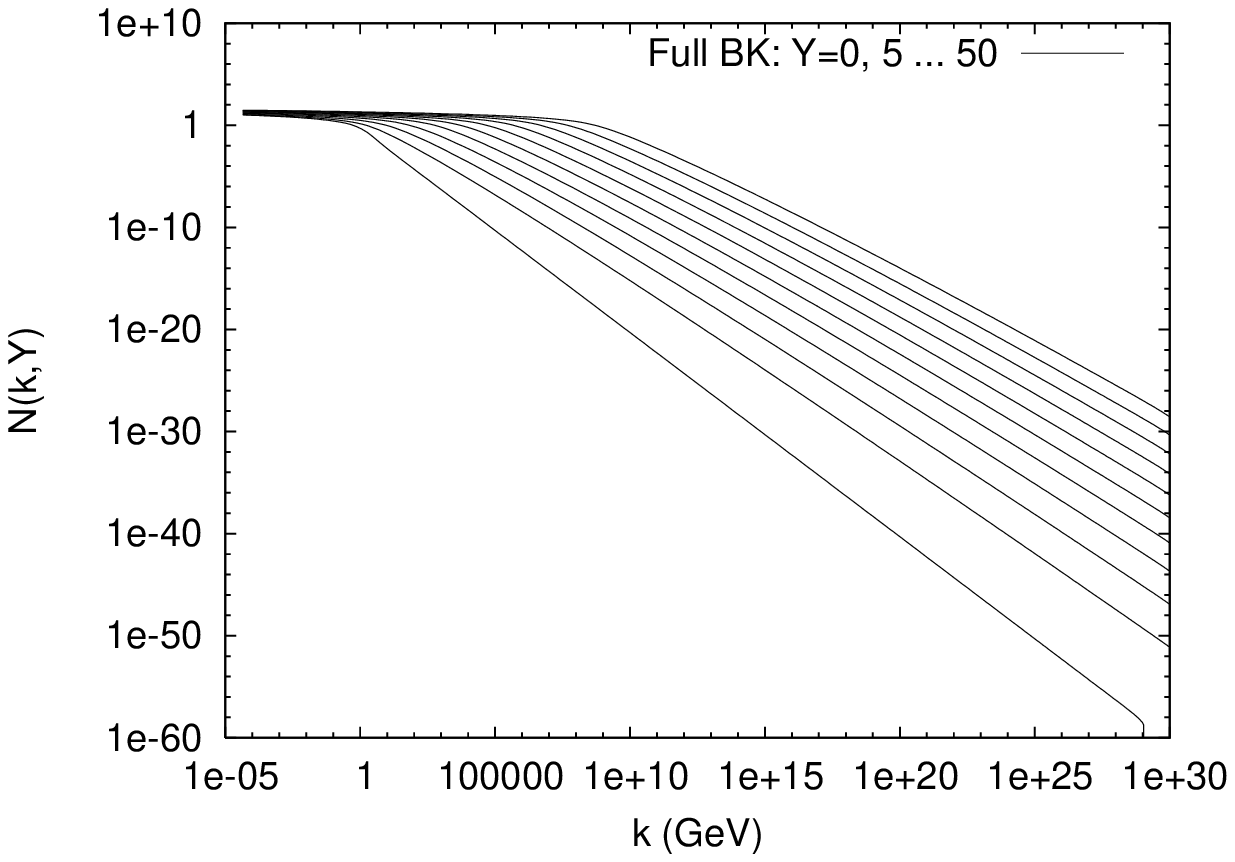,width=\wid}
\epsfig{file=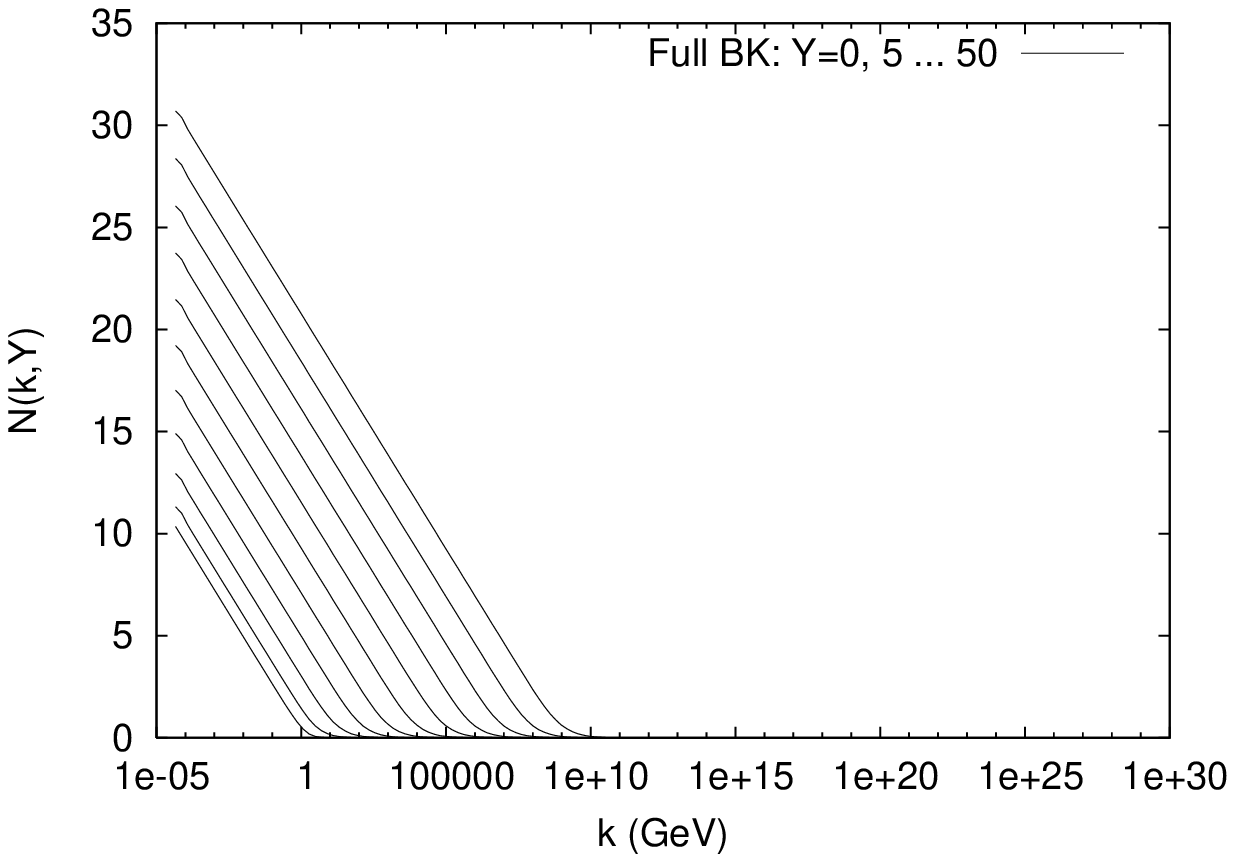,width=\wid}
\end{center}
\caption{Evolution of the MV initial condition using the full 
BK equation~(\ref{bk}), 
showing ${T}(k,Y)$ for $Y=0, 5 \dots 50$ on a logarithmic scale and a 
linear scale. The $Y=0$ curve shows the initial condition.
\label{fronts1}}
\end{figure}

Fig.\ \ref{fronts1} shows the BK evolution of the MV initial condition for rapidities up 
to $Y=50$, displaying a traveling wave behavior. The first logarithmic plot clearly 
shows the steep exponential fall-off~{(\ref{shape})} at very large momenta, 
while the second linear plot shows 
the mild $\ln(\Qs^2/k^2)$ rise towards small momenta. We have checked that
the same traveling wave front pattern persists for rapidities up to $Y=250$.

\begin{figure}[th]
\begin{center}
\epsfig{file=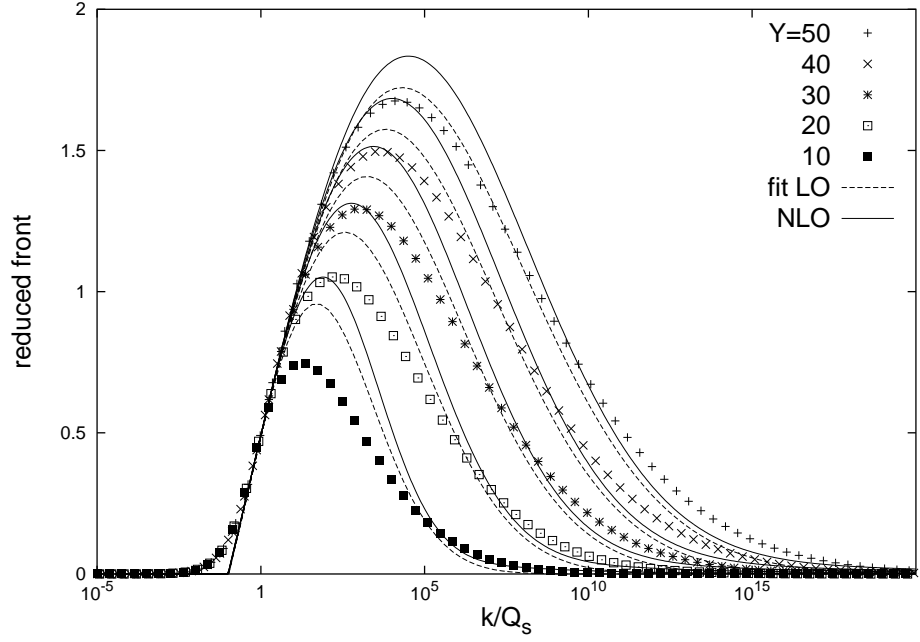,width=\wid}
\end{center}
\caption{Fit of the analytical formula for the reduced front to the result 
from the numerical calculation. Dashed line: leading order, Eq.~(\ref{frontnl}).
Full line: leading+next-to-leading order, Eq.~(\ref{eq-redfront}).
Points: result of the numerical integration for various values of the
rapidity.
\label{redfront}}
\end{figure}

To study more precisely 
the shape of the amplitude,
we plot the reduced front $\psi(k,Y)$ defined in Eq.~(\ref{defredfront}), 
which enables us to
monitor how the asymptotic traveling wave is approached.
We display the numerical results of BK evolution for different values of
the rapidity on Fig.~\ref{redfront}.
We compare them to the analytical predictions at both leading and next-to-leading
order, Eqs.~(\ref{eq-redfront}) and~(\ref{frontnl}).
Technically, we start with Eq.~(\ref{eq-redfront}), and we fit the two
constants $C_1$ and $C_2$ on the
highest rapidity data ($Y=50$): we get the values
$C_1=0.17$ and $C_2=2.9$. Then, using this determination of the parameters, 
we compute the reduced front for the 
lower rapidities, using both the leading order formula Eq.~(\ref{eq-redfront})
and the next-to-leading order prediction~(\ref{frontnl}).
It turns out that the latter does not differ very much from the former.
The agreement is not perfect, even when we use the next-to-leading order formula.
This reflects the fact that rapidities of the order of 50 are still too small
for these formulas to apply: the convergence of the asymptotic series that gives
the shape of the front is only algebraic, and hence very slow.
Actually, a better agreement in this low rapidity region
could be obtained by using our freedom
to shift the rapidity $Y\rightarrow Y+Y_0$ \cite{MP2004}.

\begin{figure}[th]
\begin{center}
\epsfig{file=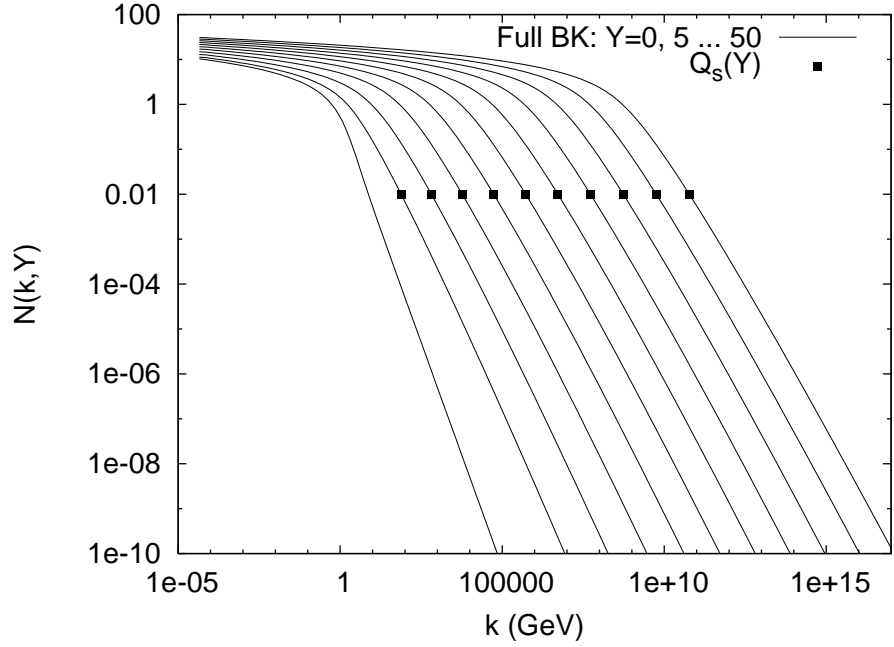,width=\wid}
\end{center}
\caption{Propagating front, with the saturation scale for $\kappa=0.01$ marked 
at each value of rapidity.
\label{frontswithqs}}
\end{figure}

We now turn to the saturation scale. Here, we define it
as the value of $k$ for which
the amplitude ${T}(k,Y)$ reaches a certain predefined value $\kappa$
(as defined in Eq.~(\ref{defpos1})). In Fig.~\ref{frontswithqs} we show 
an example of amplitudes at several rapidities, with the saturation scale marked 
as a point for each curve for the choice  $\kappa=0.01$.

\begin{figure}[th]
\begin{center}
\epsfig{file=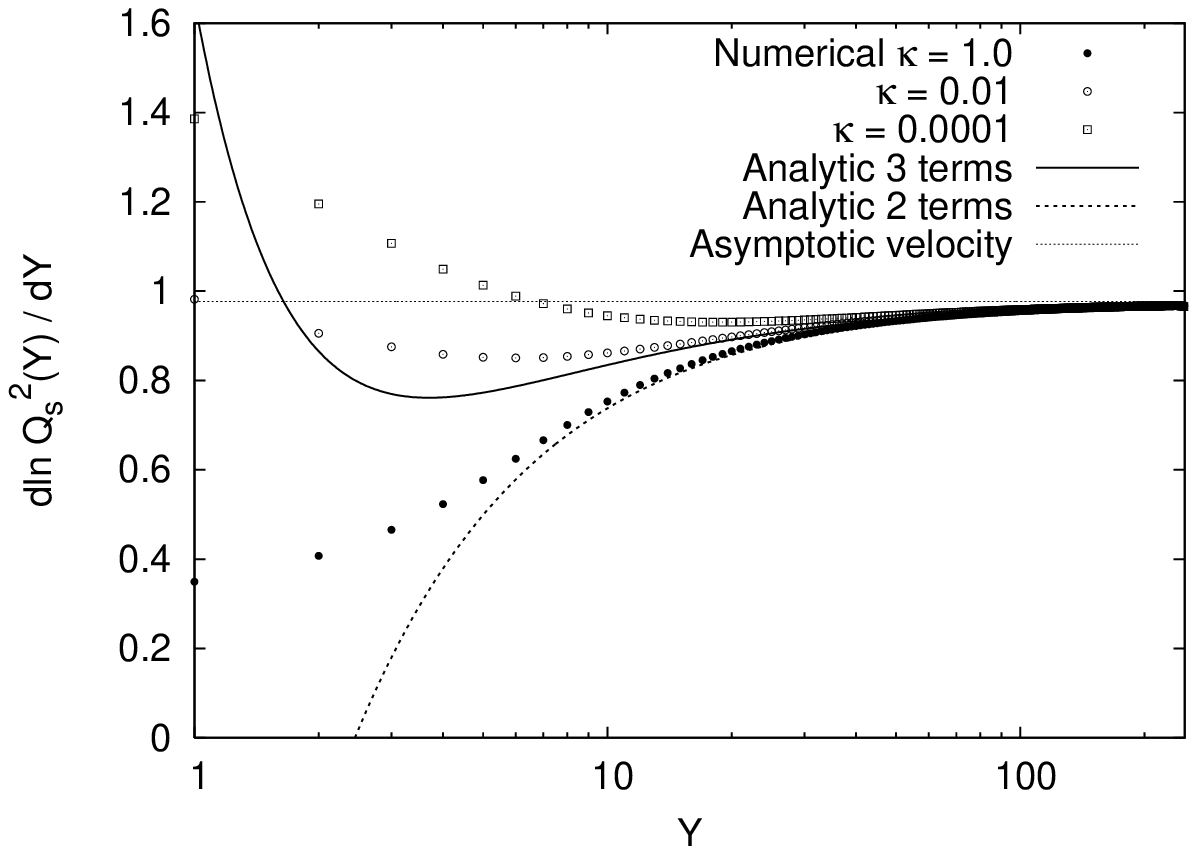,width=\wid}
\epsfig{file=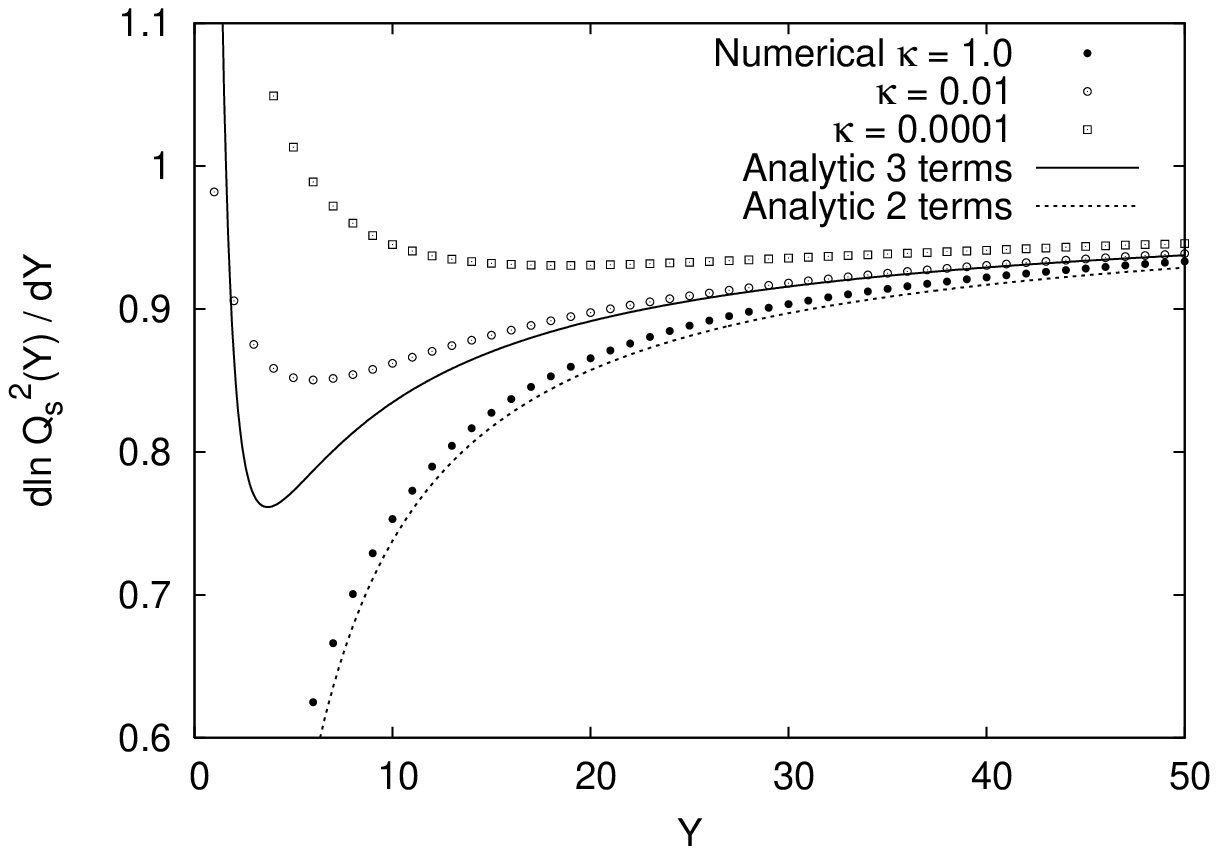,width=\wid}
\end{center}
\caption{Logarithmic derivative of the saturation scale 
$\frac{\partial \ln Q_s^2(Y)}{\partial Y}$ obtained from numerical 
simulations compared to the analytical results including two and three 
terms in the $Y$-expansion of Eq.~(\ref{satscalnl}). 
The saturation scale was obtained by tracking 
the amplitude at heights $\kappa=1.0, 0.01$ and $0.0001$. The results are 
shown for two different rapidity ranges, and on the first plot we in 
addition mark the asymptotic velocity $\as \chi(\gamma_c)/\gamma_c\approx 0.9767$.  
\label{satscale1}}
\end{figure}

Because the formation of the traveling wave front requires a large rapidity 
interval, 
the determination of the saturation scale depends on the chosen value of 
$\kappa$. We demonstrate this effect in Fig.~\ref{satscale1}, which shows 
the analytical expressions for the logarithmic derivative of the saturation 
scale (the velocity of the front) including the first two and three terms 
in the $Y$-expansion together with the numerical results for three different 
values of $\kappa$. Note that the smaller the $\kappa$-value, the longer it 
takes to reach the asymptotic limit. 
The larger the $\kappa$, the closer 
to the two-term analytic result (which needs to be closer to the asymptotic limit
to be good),
which at first sight looks quite intriguing \cite{GB2004}. 
Expressed in another way, different parts of the wave front travel with 
different velocities, which only become equal in the asymptotic $Y$ limit.

It is clear, however, that the analytical results
agrees very well with 
the numerical results for large $Y$, where they tend to the constant velocity 
$\as\chi(\gamma_0)/\gamma_0$, corresponding to the exponential increase of the 
saturation scale. For smaller rapidities, 
we see clearly the effect of subasymptotic effects on the numerical calculation.
In particular, the dependence of the numerical
results on the very definition of the saturation scale is manifest.

\begin{figure}[th]
\begin{center}
\epsfig{file=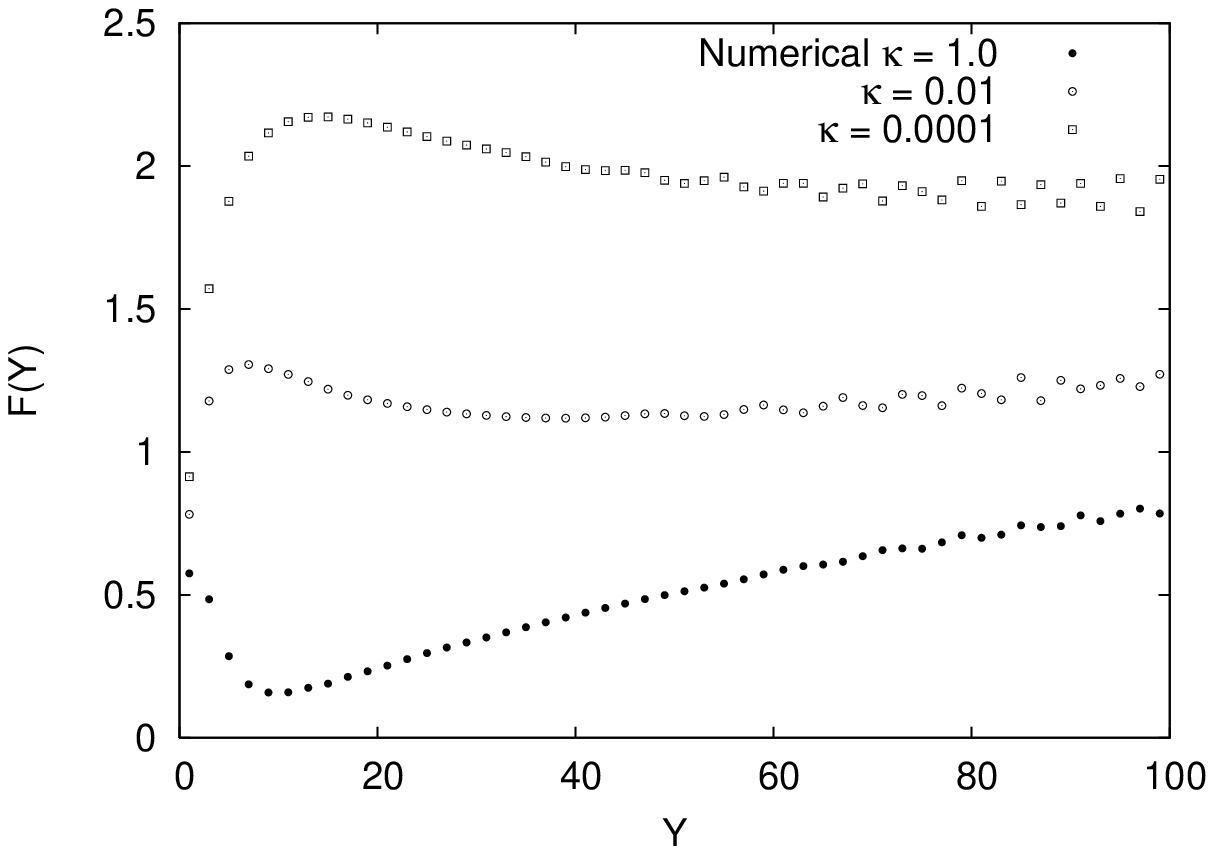,width=\wid}
\end{center}
\caption{${ F}(Y)$ (Eq.~(\ref{eqF})) for $\kappa=1.0, 0.01$ and $0.0001$.
\label{red_dqs1}}
\end{figure}

\begin{figure}[th]
\begin{center}
\epsfig{file=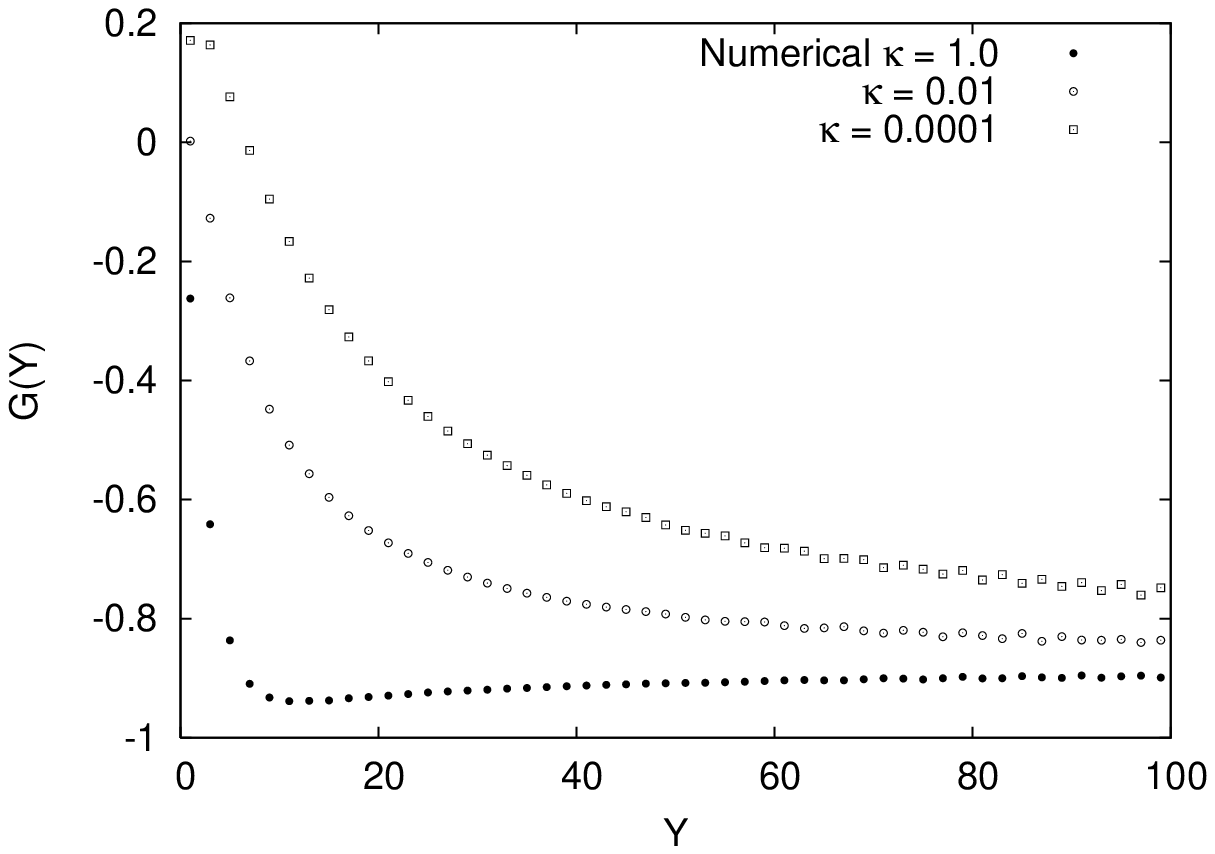,width=\wid}
\end{center}
\caption{${ G}(Y)$ (Eq.~(\ref{eqG})) for $\kappa=1.0, 0.01$ and $0.0001$.
\label{red_dqs2}}
\end{figure}

In order to study 
the agreement of the numerical calculation with  Eq.~(\ref{satscalnl}) 
at subleading level
in more detail we define the two quantities
\begin{align}
\label{eqF}{ F}(Y) &= Y^{3/2}\frac{2\gc^2}{3}\sqrt{\frac{\bar\alpha\xggc}{2\pi}}
\left( \frac{\partial \ln \Q}{\partial Y} 
-\frac{\bar\alpha \xc}{\gc} + \frac{3}{2\gc} \frac{1}{Y}\right)
\\
\label{eqG}{ G}(Y) &= Y \frac{2\gc}{3}
\left( \frac{\partial \ln \Q}{\partial Y} 
-\frac{\bar\alpha \xc}{\gc}\right)\ ,
\end{align}
which, according to the analytical formulas asymptotically behave as 
${ F}(Y) \to 1 + {\cal O}(1/\sqrt Y)$ and
${ G}(Y) \to -1 + {\cal O}(1/\sqrt Y)$ 
(cf. the corresponding expression for the saturation scale in Eq.\ (\ref{satscalnl})). 
In the absence of a third term in Eq.~(\ref{satscalnl}), 
${ F}(Y)$ would instead be close to zero, 
while ${ G}(Y)$ would be close to $-1$. These quantities, extracted from the 
numerical determination of the saturation scale, are plotted in Figs.\ \ref{red_dqs1} 
and \ref{red_dqs2}. These plots clearly indicate that the third ${\cal O}(1/\sqrt Y)$-term 
in the saturation scale is present until rather large rapidities. Furthermore, its 
influence is smaller when the saturation scale is determined ``higher up'' on the 
front, where the evolution more quickly approaches asymptotics.

\begin{figure}[th]
\begin{center}
\epsfig{file=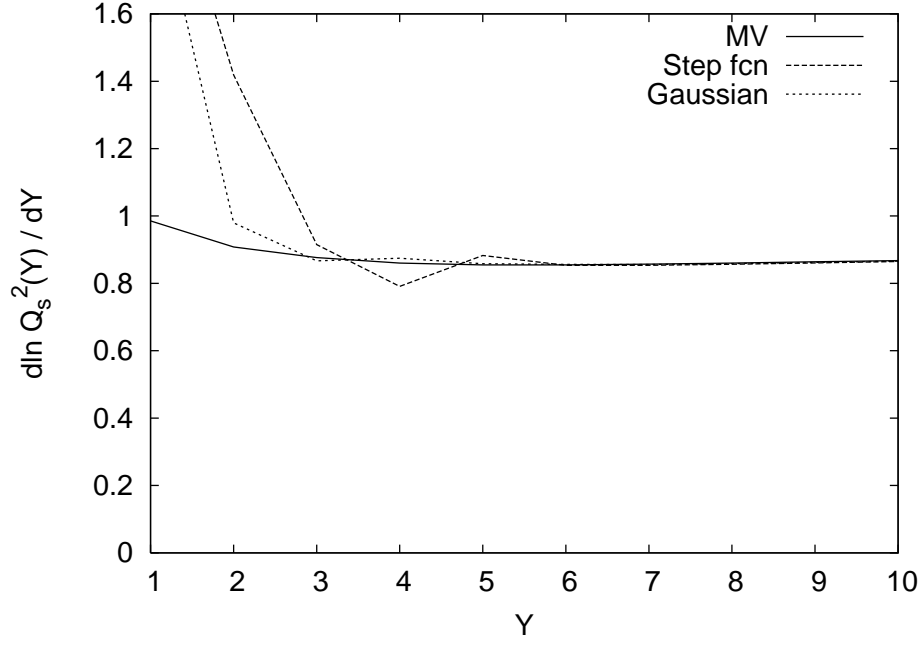,width=\wid}
\end{center}
\caption{Front velocity for three different initial conditions:
McLerran-Venugopalan \cite{MV}, a 
step function, and a Gaussian. 
\label{satscaleic}}
\end{figure}

\begin{figure}[th]
\begin{center}
\epsfig{file=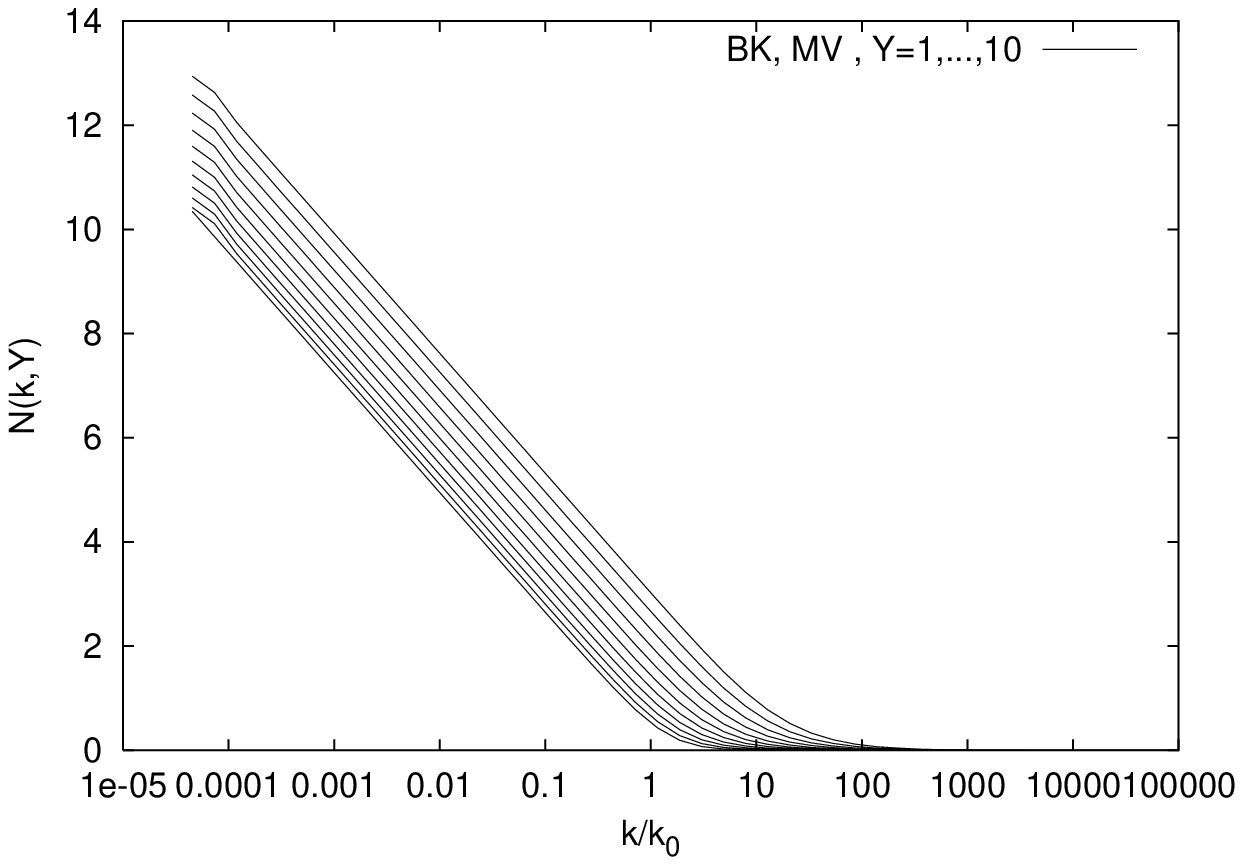,width=0.32\columnwidth}
\epsfig{file=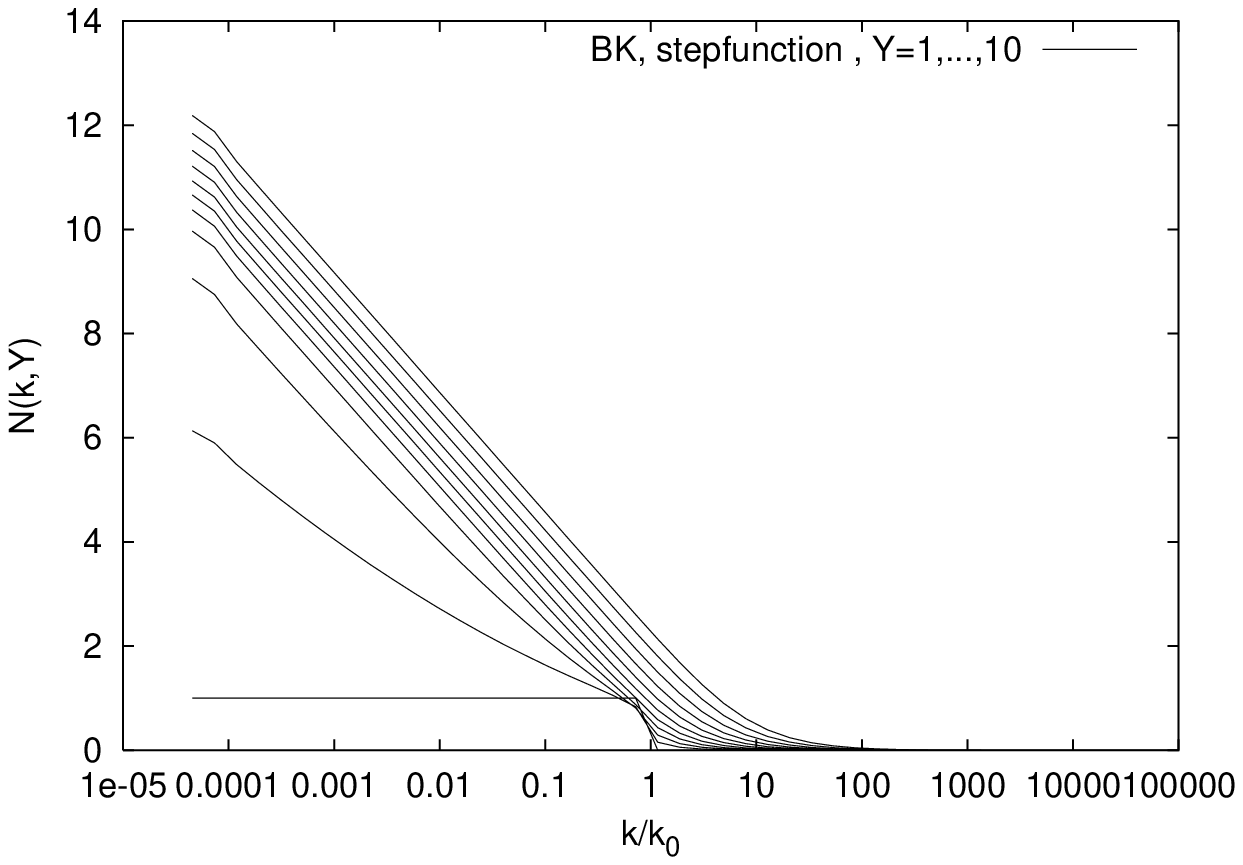,width=0.32\columnwidth}
\epsfig{file=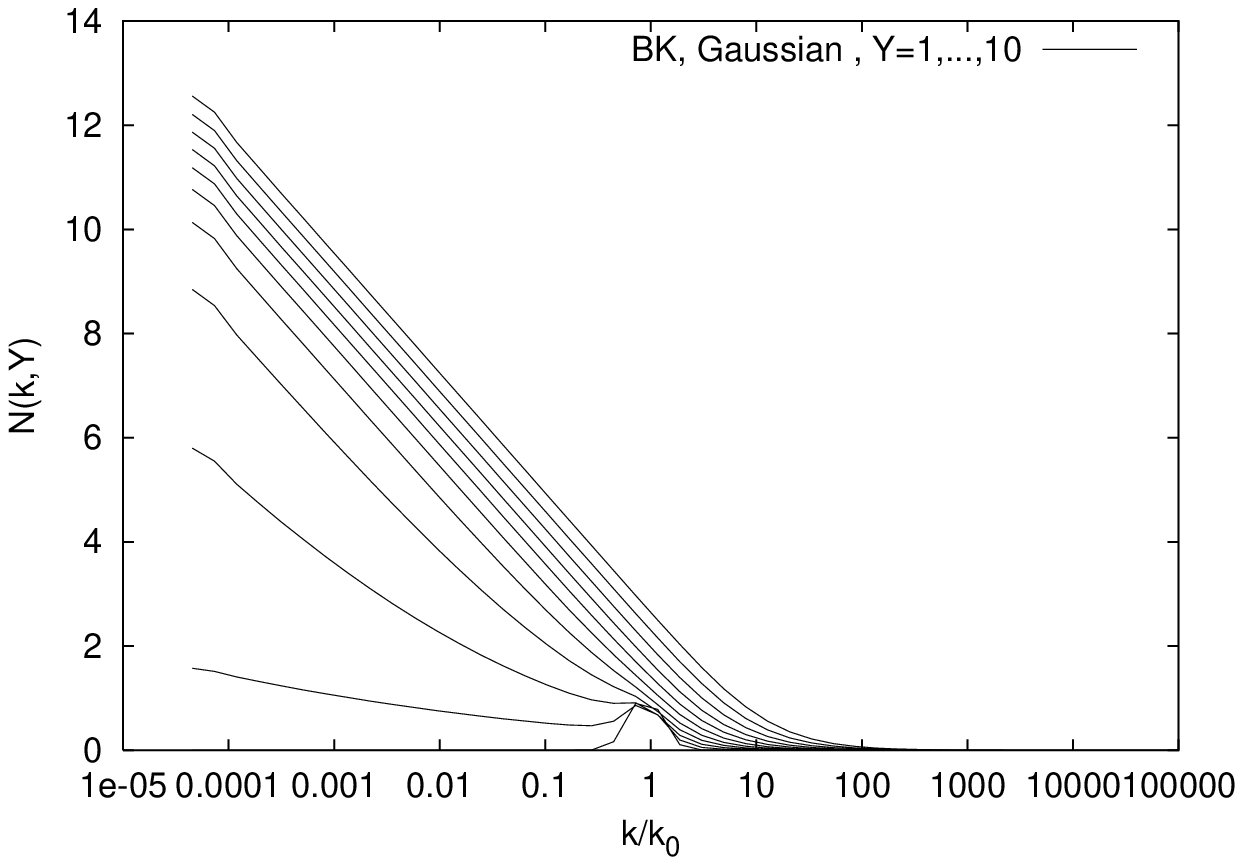,width=0.32\columnwidth}
\end{center}
\caption{Wave fronts for three different initial conditions:\ 
McLerran-Venugopalan, a step 
function, and a Gaussian (from the left to the right). 
\label{icfronts}}
\end{figure}

The saturation scale has a non-universal dependence on the initial condition, 
which is $Y$-suppressed compared to the three terms 
in the expression (\ref{satscalnl}).
To study this dependence we determine the velocity of the evolution from three 
different initial conditions, the MV model, a step function $\Theta(L)$, and a 
Gaussian centered at $k=k_0$. The result, see Fig.~\ref{satscaleic}, shows 
that the three curves converge to the same propagation velocity within a
few units in rapidity. 
The corresponding wave fronts are shown in Fig.~\ref{icfronts}.

\begin{figure}[th]
\begin{center}
\epsfig{file=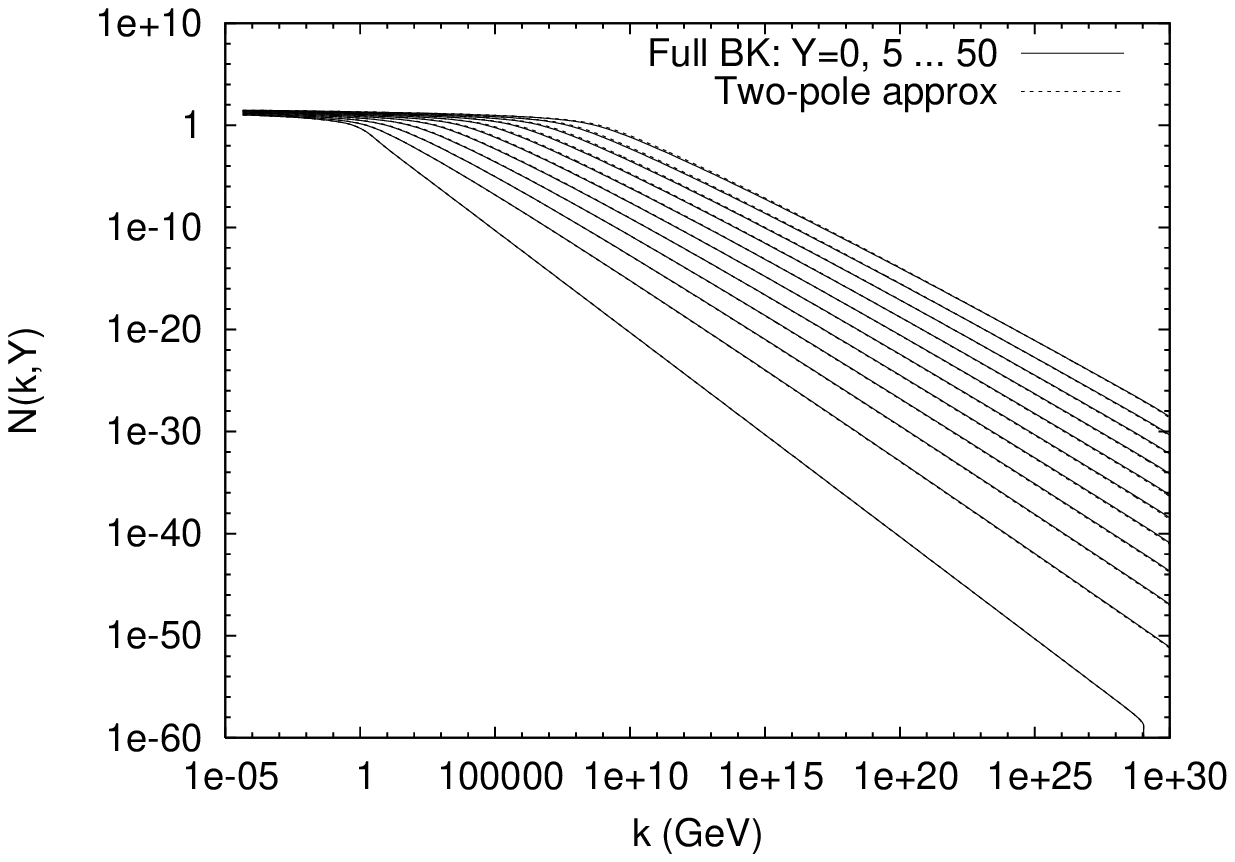,width=\wid}
\epsfig{file=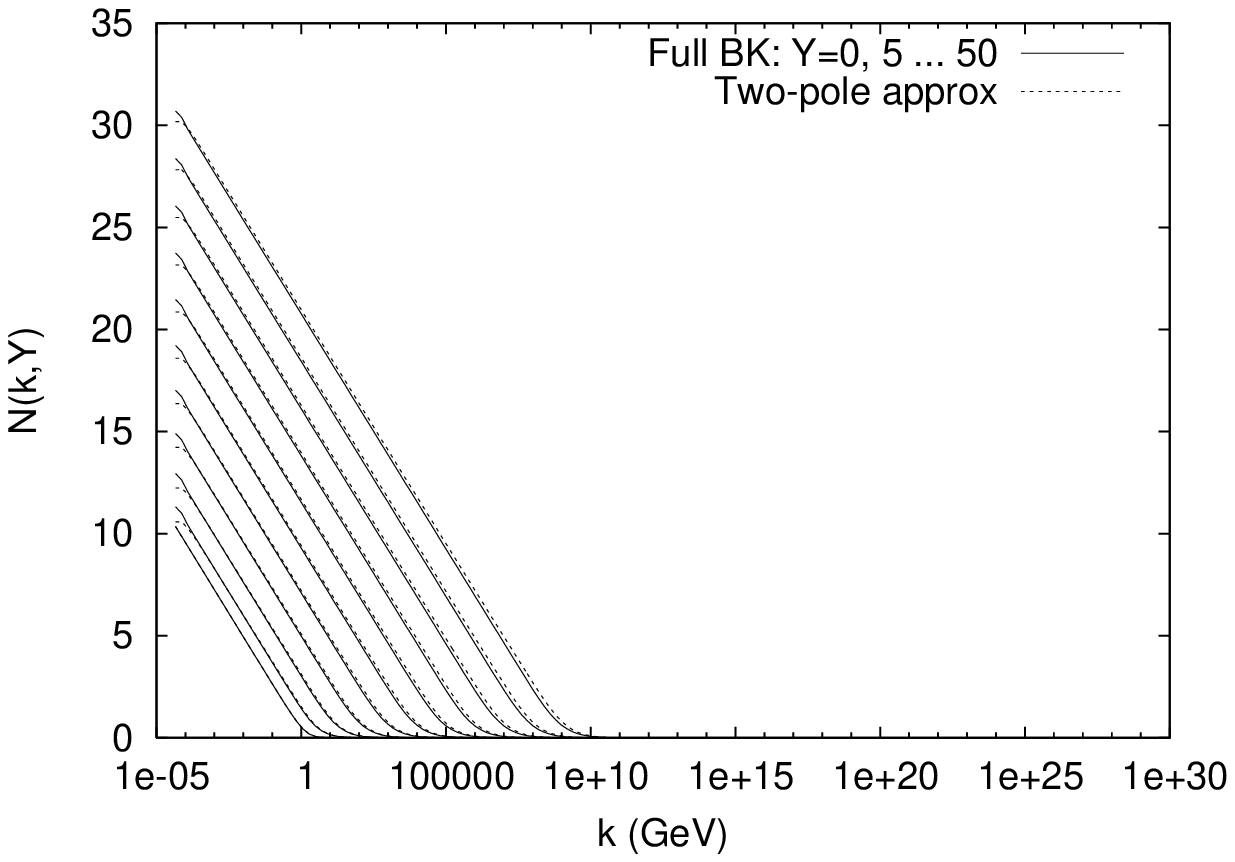,width=\wid}
\end{center}
\caption{Comparison between evolution by the full BK kernel, 
Eq.\ (\protect\ref{numBKkernel}), and the two-pole approximation 
to the kernel, Eq.\ (\protect\ref{num2pkernel}).
\label{fronts3}}
\end{figure}

\begin{figure}[th]
\begin{center}
\epsfig{file=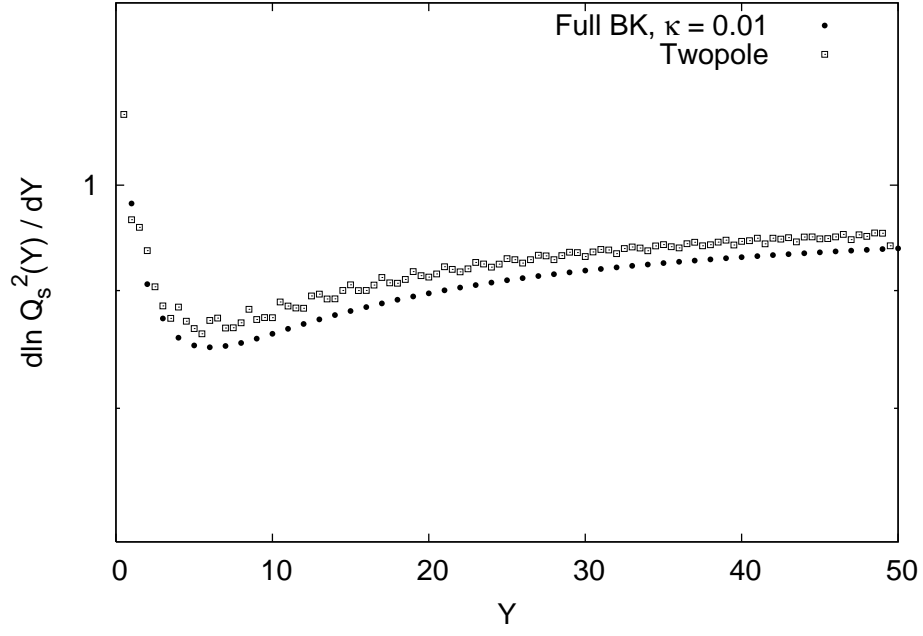,width=\wid}
\end{center}
\caption{Comparison of the logarithmic derivative of the saturation 
scale $\frac{\partial \ln Q_s^2(Y)}{\partial Y}$ as obtained from 
evolution using full BK and using the two-pole approximation of the kernel.
\label{satscale2p}} 
\end{figure}

Finally,
we would also like to know to what extent the two-pole approximation of the 
BFKL kernel, represented in momentum space by Eq.\ (\ref{num2pkernel}), 
captures the essentials of the evolution. Fig.\ \ref{fronts3} shows a 
comparison between the amplitudes obtained from the two kernels. 
Evidently they give very similar results. 
We also compare the velocities in Fig.\ \ref{satscale2p}.

A few comments are in order. There is an obvious 
difference between the solutions
of the FKPP equation, which corresponds to the diffusive 
approximation~(\ref{chiexpansion}) and those of the BK equation.
While the FKPP waves are bounded by a constant for small $k\ll Q_s$,
the BK amplitude diverges like $\ln Q_s^2/k^2$, as seen on
Fig.~\ref{fronts1}. This stems from the nonlocality of
the full BFKL kernel, and can most easily be understood on the
simplified form of that kernel in 
Eq.~(\ref{num2pkernel}). Indeed, $T(L^\prime)=\mbox{constant}$ does not diagonalize
the kernel in the region $L\rightarrow -\infty$.
Instead, one iteration of the kernel 
generates a factor $L_s-L=\ln(Q_s^2/k^2)$ 
from the first term in Eq.~(\ref{num2pkernel}).
The same phenomenon occurs in the large $k$ region, in which the
behavior $e^{-\gamma_0 L}$ gets enhanced by powers of $L$.
However, this does not influence the properties of
the traveling waves that we are studying here. 
The reason is that these extra factors are subleading with respect
to the powers of $k^2$ that determine the asymptotic behavior of the
traveling wave.
Our numerical
study shows an excellent quantitative agreement 
with the analytical solutions~(\ref{satscalnl})-(\ref{frontnl}).

 We note that it is also possible to extract the saturation scale from the 
numerical results by fitting a functional form $\ln(Q_s^2/k^2)$ to the numerical 
curves for small $k$~\cite{GB2004}.

\begin{figure}
\begin{center}
\epsfig{file=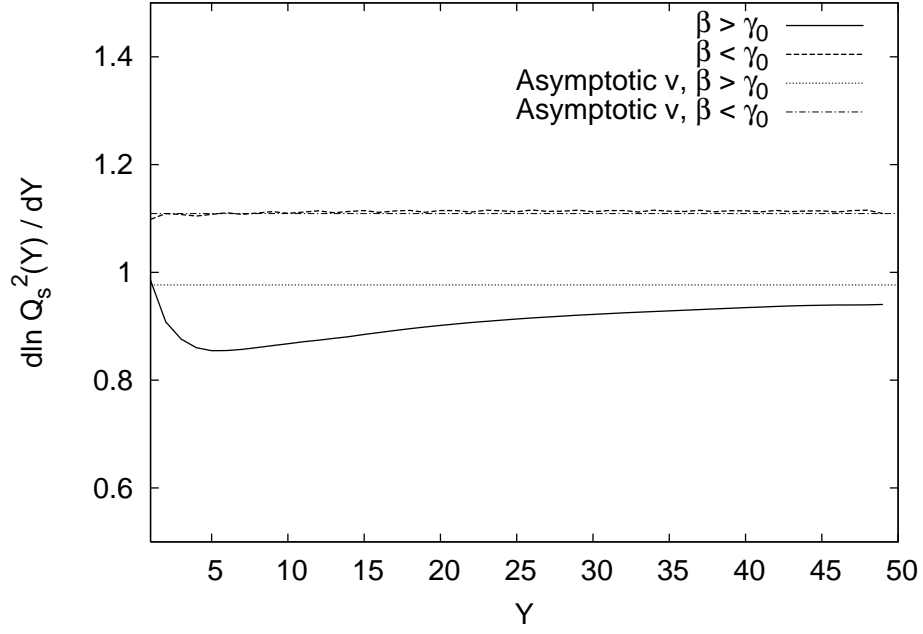,width=\wid}
\end{center}
\caption{\label{dqs_notsteep}
Comparison of the 
mean field
rapidity evolutions of two different initial conditions: 
McLerran-Venugopalan (Eq.~(\ref{MV}), full line),
and an initial condition which behaves like $1/k$ for large $k$ (dashed line).
The plot shows the logarithmic derivative of the saturation 
scale $\frac{\partial \ln Q_s^2(Y)}{\partial Y}$ in both cases.
The straight horizontal lines are the large rapidity asymptotics
(first term in Eq.~(\ref{satscalnl}) and Eq.~(\ref{scritical}) resp.)
}
\end{figure}

\begin{figure}
\begin{center}
\epsfig{file=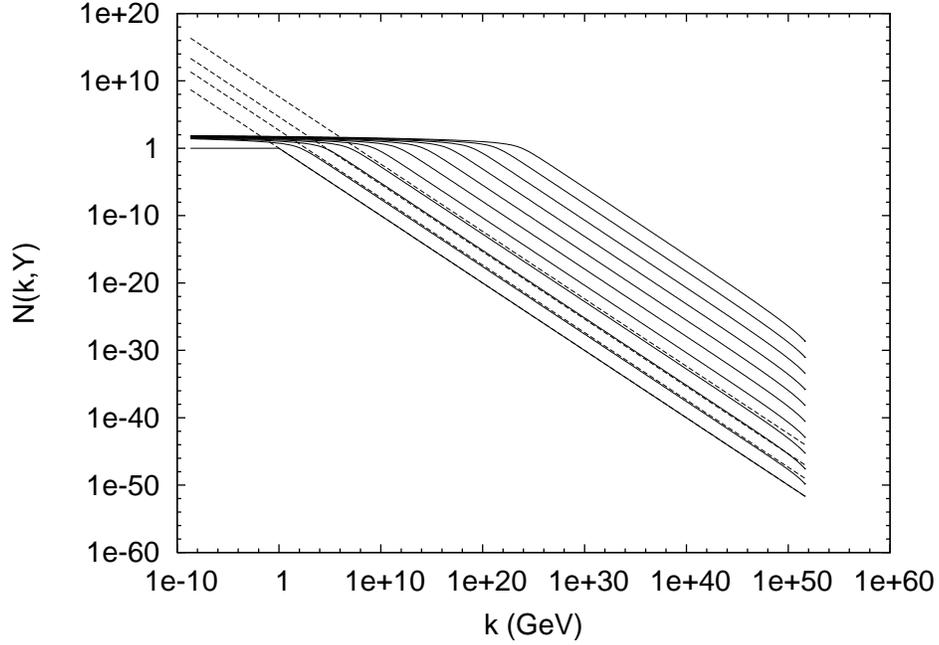,width=\wid}
\end{center}
\caption{\label{fronts_notsteep}
Evolution of the shape of the front upon rapidity evolution,
starting from an initial condition that behaves like $1/k$ at large $k$.
The rapidity interval ranges from $Y=0$ (initial condition) to $Y=100$ (from the left
to the right). The dotted lines represent $C/k$.
}
\end{figure}

So far, we have focused on a physical initial condition. We considered the MV
model, but any initial condition that satisfies color transparency at large $k^2$
would satisfy the condition $\beta>\gamma_0$, and hence, the asymptotic saturation scale
would obey Eq.~(\ref{satscalnl}) and the asymptotic shape of the front at large $Y$
and large $k$ would be $T(k)\sim 1/k^{2\gamma_0}$. 
We wish however to test further the theoretical framework outlined 
in Sec.~\ref{sec:general} by picking the non-physical initial condition $T(k)\sim 1/k$,
in which case $\beta=\frac12<\gamma_0$.
In this case, the asymptotic saturation scale should be
\begin{equation}
\ln Q_s^2(Y)/dY=\bar\alpha_s\chi(\beta)/\beta\simeq 1.1090\cdots
\label{scritical}
\end{equation} 
and the shape of the front
should be conserved.
This is precisely what is observed in the results of the numerical calculation,
see Figs.~\ref{dqs_notsteep} and~\ref{fronts_notsteep}, 
in complete agreement with the theoretical expectations
in Sec.~\ref{sec:general}.
We note however that
our results clearly disagree with the claims made in Ref.~\cite{AAMSW},
where it was found that the asymptotic amplitude
goes like $T(k)\sim 1/k^{2\gamma_0}$ at large $k$ in all cases, $\beta>\gamma_0$ and
$\beta<\gamma_0$.


\section{\label{sec_stochastic}
Beyond the mean field: universal features from a toy model}

In Sec.~\ref{sec_setup}, we identified the universality class of high energy QCD
as that of a stochastic partial differential equation, the sFKPP equation.
We studied in detail its mean field approximation in Sec.~\ref{sec_meanfield}.
We now aim at investigating the effects induced by the stochastic terms
neglected in that approximation.

For this purpose, we will propose a simple toy model that is inspired
by the QCD dipole model, and that belongs to the same universality
class.
The advantage of studying a toy model are numerous:\ First, all universal
results can be checked and illustrated in a simple and physical way, leaving out
the irrelevant complications of QCD.
Second, many qualitative features, that also hold in the case of QCD, 
can be investigated. For example, 
we are going to study in detail
the relevance of the mean field
approximation that was investigated in Sec.~\ref{sec_meanfield}.


\subsection{General results on the sFKPP equation}
\label{generalresults}

Let us start by reviewing the most recent general results that
have been obtained regarding universal features 
of the solutions of the sFKPP equation (or equivalently of the Reggeon field theory
equation~(\ref{RFT})).
We will then formulate our toy model in Sec.~\ref{subsec_toy} and illustrate its
universal properties outlined here in a numerical solution in Sec.~\ref{subsec_numsol}.

The mean shape of the front in the region where $1/N\ll u\ll 1$ 
was found to be \cite{BD}
\begin{equation}
u(x,t)\sim \frac{\ln N}{\pi\gamma_0} 
\sin\left(\frac{\pi\gamma_0(x-X_t)}{\ln N}\right)e^{-\gamma_0(x-X_t)}\ ,
\label{shapeBD}
\end{equation}
where $\gamma_0$ minimizes the function $v(\gamma)$ 
of Eq.~(\ref{vgamma}).
The large time velocity of the wave front was shown to be
\begin{equation}
v=v(\gamma_0)-\frac{\pi^2\gamma_0^2 v^{\prime\prime}(\gamma_0)}
{2\ln^2 N}\ .
\label{vBD}
\end{equation}
This is expected to be the first terms of an asymptotic expansion 
for large $N$.
Terms of relative order $1/\ln N$ are neglected.
The scaling of the second moment of the front position is also
known from numerical simulations of different models~\cite{BD}
\begin{equation}
\sigma^2=\langle X_t^2\rangle-\langle X_t\rangle^2
\propto\frac{t}{\ln^3 N}\ .
\label{sigma2Xt}
\end{equation} 
 From Eq.~(\ref{sigma2Xt}) and~(\ref{shapeBD}), 
we obtain easily the scaling form of the average of $u$, which
corresponds to the physical amplitude in the case of QCD \cite{IMM}:
\begin{equation}
\langle u(x,t)\rangle=u\left(\frac{x-\langle X_t\rangle}{\sqrt{
t/\ln^3 N}}\right)\ ,
\label{scaling}
\end{equation}
up to ${\mathcal O}(1/\sqrt{t})$ corrections. 

While these results rely mostly on conjectures written down after
a numerical study of discrete models \cite{BD},
they are believed to have a general validity.
Indeed, they are supported both by appealing physical arguments, that
we will develop below, and by
accurate numerical
calculations within different specific models.

In the following, we will propose a toy model that is very similar to QCD
and confront the numerical solutions with the above-mentioned 
analytical predictions.


\subsection{\label{subsec_toy}A simple toy model}

We consider a set of particles 
on a one-dimensional lattice,  whose
sites are labeled by the coordinate $x$ and separated by the distance $\Delta x$.
These particles may
jump from one position to the nearest one on the left or on the right
with probability $p_L$ and $p_R$ respectively, and they can multiply with
probability $\lambda$. These elementary
processes replace the linear diffusive growth of 
partons as described by the BFKL evolution.
In order to enforce a limit on the number of particles on the different sites,
we impose that each of the $n(x,t)$ particles piled up at site $x$ at time $t$
may die with a probability $\lambda n(x,t)/N$. 
This is a way of implementing the equivalent of parton saturation. 
These rules completely define the model.

This setup is typical of a ``reaction-diffusion'' process.
It could realistically represent, for example, bacterial growth, but
more generally it represents a wide universality class, to which
high energy QCD also belongs, as was explained in Sec.~II.

It is straightforward to convert the evolution laws 
stated above in an equation for
the number of particles $n(x,t)$ on site $x$.
Between times $t$ and $t+\Delta t$, there are $n_L(x)$ particles which
jump to the left, $n_R(x)$ particles which jump to the right, $n_+(x)$
which multiply, and $n_-(x)$ which disappear. Thus the variation
in the particle number reads
\begin{equation}
n(x,{t\!+\!\Delta t})-n(x,t)=-n_L(x)-n_R(x)-n_-(x)+n_+(x)
+n_L(x\!+\!\Delta x)+n_R(x\!-\!\Delta x)|_{t\rightarrow t+\Delta t}\ .
\label{stochatoy}
\end{equation}
>From the rules edicted above, each configuration has
a multinomial distribution
\begin{multline}
P(\{n_L(x),n_R(x),n_+(x),n_-(x)\})
=
\frac{(n(x,t))!}
{(n_L(x))! (n_R(x))! (n_+(x))! (n_-(x))!
(\Delta n(x,t))!}
\\
\times p_L^{n_L}p_R^{n_R}\lambda^{n_+}
(\lambda n(x,t)/N)^{n_-}
(1\!-\!p_L\!-\!p_R\!-\!\lambda\!-\!\lambda n(x,t)/N)
^{\Delta n(x,t)}\ ,
\label{stocharule}
\end{multline}
where $\Delta n(x,t)=n(x,t)\!-\!n_L(x)\!-\!n_R(x)\!-\!n_+(x)\!-\!n_-(x)$.

To get a clearer picture of the evolution, one may compute
the mean change in the particle number in one time step
from Eqs.~(\ref{stochatoy}) and~(\ref{stocharule}):
\begin{equation}
\langle n(x,{t+\Delta t})|\{n(x,t)\}\rangle=n(x,t)(1-p_L-p_R-\lambda n(x,t)/N)
+p_L n(x\!+\!\Delta x,t)
+p_R n(x\!-\!\Delta x,t)+\lambda n(x,t) \ .
\label{meantoy}
\end{equation}
Obviously, there are two fixed points of the mean evolution: $n(x,t)=0$ and
$n(x,t)=N$. The latter is the maximum number of particles
allowed on each site, on the average. Let us introduce 
the fractional occupation number $u(x,t)=n(x,t)/N$.
The mean evolution of $u(x,t)$ in one step in time reads
\begin{equation}
\left\langle u(x,{t+\Delta t})|\{u(x,t)\}\right\rangle=u(x,t)
+p_L [u(x\!+\!\Delta x,t)-u(x,t)]
+p_R [u(x\!-\!\Delta x,t)-u(x,t)]+\lambda u(x,t)(1-u(x,t)) \ .
\label{avu}
\end{equation}
Taking the average of both sides of this equation, one sees that it is not
a closed equation for $\langle u\rangle$.
It gets closed only after 
a mean field approximation, obtained through the replacement
$u \rightarrow \langle u\rangle$.

The phase velocity of a wave of wave number $\gamma$ may be obtained 
from Eq.~(\ref{avu}) by
computing the mean evolution of the plane wave  $e^{-\gamma(x-vt)}$
in one linear evolution step
\begin{equation}
v(\gamma)=\frac{1}{\gamma\Delta t}
\ln\left(1+\lambda+p_L (e^{-\gamma\Delta x}-1)+p_R (e^{\gamma\Delta x}-1)
\right)\ .
\label{vgammatoy}
\end{equation}
This function 
is the equivalent of $v(\gamma)$ in Eq.~(\ref{vgamma}) in the present context.
It is enough to compute the corresponding values of $\gamma_0$, $v(\gamma_0)$,
$v^{\prime\prime}(\gamma_0)$
and to replace them in
Eqs.~(\ref{shapeBD}),(\ref{vBD})
to get the predictions for the asymptotic shape and velocity of the front.

For the purpose of our numerical study, we choose
\begin{equation}
\Delta x=\Delta t=1\ ,\ \ p_L=p_R=0.1\ ,\ \ \lambda=0.2\ ,
\end{equation}
which lead to the following parameters:
\begin{equation}
\gamma_0=1.35219\ ,\ \ v(\gamma_0)=0.255386\ ,\ \
v^{\prime\prime}(\gamma_0)=0.167721\ .
\end{equation}
We pick an initial condition that consists of $N$ particles
sitting at $x=0$ at time $t=0$ ($u(x\!=\!0,t\!=\!0)=1$).
The position of the front $X_t$ is chosen to be
defined by the prescription in Eq.~(\ref{defpos2}).

\begin{table}
\begin{center}
\begin{tabular}{|c|c|}
\hline
QCD & Toy model\\
\hline
$\bar\alpha Y$ & $t$\\
$v(\gamma)=\chi(\gamma)/\gamma$ & $v(\gamma)=$ Eq.~(\ref{vgammatoy}) \\
$L=\ln k^2/\Lambda^2$ & $x$ \\
$L_s=\ln Q_s^2/\Lambda^2$ & $X_t$\\
$T(k,Y)$ & $u(x,t)$\\
$\alpha_s$ & $1/\sqrt{N}$\\
\hline
\end{tabular}%
\end{center}%
\caption{\label{dictionary}Dictionary between QCD and the toy model settings.}
\end{table}

\begin{figure}
\begin{center}
\epsfig{file=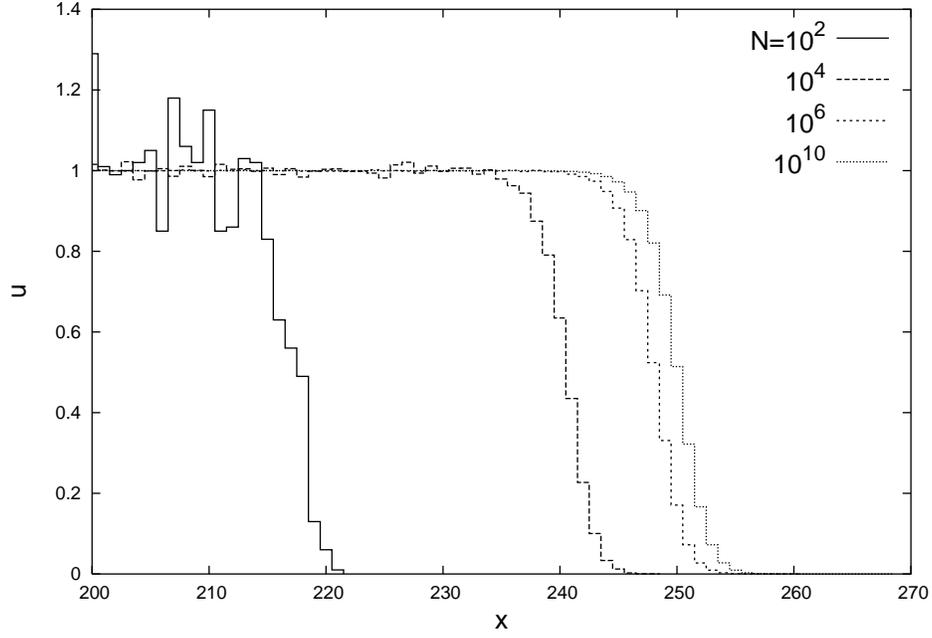,width=\wid}
\end{center}
\caption{\label{front1}Numerical integration of the toy model defined
in Eqs.~(\ref{stochatoy},\ref{stocharule})
over 1000 units of time for different values of $N$.}
\end{figure}

We evolve that initial condition for different values of $N$.
The result is shown in Fig.~\ref{front1} for $t=1000$.
It is a front with fluctuations around a mean steady shape.
Recall that these fluctuations are statistical fluctuations 
in the particle numbers. We see that their magnitude is consistent with 
the expectation 
\begin{equation}
\delta u\sim\sqrt{u/N}\ .
\label{fluctbin}
\end{equation}

In view of these results, and in order to be able to go to larger
values of $N$, we choose to replace the stochastic evolution~(\ref{stochatoy}) 
by its mean field approximation~(\ref{meantoy}) in the bins in which the number of
particles is larger than say $10^4$, together with an appropriate
boundary condition on the interface between the stochastic
and deterministic calculation. 
According to Eq.~(\ref{fluctbin}),
this amounts to neglecting 
effects of order 1\% in the concerned bins, which we do not consider here and
which, empirically, have anyway no sizable influence neither 
on the form nor on the position 
of the front. Such a hybrid method has been considered before for reaction-diffusion
models \cite{Moro}.

We may even try to simplify more, keeping the stochasticity only in the foremost
bin, along the lines of Ref.~\cite{BD}.
At each new time step, the non-empty bins are updated using a pure
mean field approximation. A new bin is filled immediately on the right of the
rightmost nonempty site, with
a number of particles given by a Poisson law of parameter 
\begin{equation}
\theta =N\times\left\langle u(x,{t+\Delta t})|\{u(x,t)\}\right\rangle\ .
\label{poisson}
\end{equation}
where $\left\langle u(x,{t+\Delta t})|\{u(x,t)\}\right\rangle$
is given in Eq.~(\ref{avu}).
We will also use this simpler model, to get a feeling on how much
the solutions depend on the details of the model.
For our purpose, it is important to have such indications, since eventually,
we would like to draw conclusions for QCD from our toy model.


\subsection{\label{subsec_numsol}
Time dependence of the position of the front}

Starting from our localized initial condition, 
the first stages of the evolution consist in
a diffusive spreading and multiplication of the particles,
with a gradual filling of the nearby states. Indeed:
\begin{equation}
\begin{split}
\langle u(x=0,\Delta t)\rangle=1\ \ \mbox{and}\ \
\langle u(x=1,\Delta t)\rangle=p_R\ .
\end{split}
\end{equation}
As $p_R={\cal O}(1)$ the occupation number of all non-empty sites is
very large after this step, and the mean field approximation $u=\langle u\rangle$ 
is certainly justified. This argument applies
also for the few successive steps.
Thus, as the mean field is justified, we expect that 
an exponentially decaying wave front be gradually built
\begin{equation}
u(x,t)\sim e^{-\gamma_0(x-X_t)};
\label{front}
\end{equation}
see Eq.~(\ref{shape}). In these early stages,
all mean field results can be taken over from Sec.~III, with the appropriate
replacements listed in Tab.~\ref{dictionary}.

At this point, a comment is in order. The initial condition that we have taken
is similar to the McLerran-Venugopalan model considered above, which is certainly
a good representation of a nucleus or nucleon at very low rapidities.
However, in the case of a purely perturbative target, like a single dipole,
the equivalent
relevant initial condition for our toy model
would rather be $u(x=0,t=0)=1/N$.
Our argument above relies on the large occupation numbers and
cannot apply here.
However, in this case, the initial stages of the evolution
consist in building a dense state around $x=0$, with a steep tail in the direction of
$x>0$, steeper than the critical front profile $e^{-\gamma_0 x}$.
This process requires a time of order $\ln N$.
Subsequently, the steep profile is converted into the critical one 
through a mean field evolution.

>From Eq.~(\ref{diffusiontmf}) with the appropriate
identifications listed in Tab.~\ref{dictionary}, it takes a time of order
\begin{equation}
t\sim \frac{(x-X_t)^2}{2 \gamma_0 v''(\gamma_0)}
\label{diffusiont}
\end{equation}
for the profile~(\ref{front}) to diffuse
within the distance $x-X_t$ from the bulk of the front.
During that time, the front velocity is given by
\begin{equation}
v(t)=\frac{X_t-X_{t-\Delta t}}{\Delta t}=v(\gamma_0)-\frac{3}{2\gamma_0 t}\ ,
\label{velocityt}
\end{equation}
where we have kept only the leading correction in $t$ from Eq.~(\ref{satscalnl}).
The deterministic construction of the front
may last until the exponential shape has reached the point
where $u\sim 1/N$, where it has to stop because of discreteness of the 
number of particles: $u(x,t)=0,1/N,2/N\cdots$, and these small discrete
numbers can definitely not be interpolated by an exponential 
of the form~(\ref{front}).
>From Eq.~(\ref{diffusiont}) and from the shape of the
asymptotic front Eq.~(\ref{front}), we see that the latter process stops
after the time
\begin{equation}
t_{diff}= c\frac{(\ln N/\gamma_0)^2}
{2\gamma_0 v^{\prime\prime}(\gamma_0)} 
\label{time}
\end{equation}
at which the velocity of the front is, from Eq.~(\ref{velocityt}):
\begin{equation}
v(t_{diff})=v(\gamma_0)
-\frac{3\gamma_0^2 v^{\prime\prime}(\gamma_0)}{c\ln^2 N}\ .
\label{velocityBD}
\end{equation}
If there were no fluctuations in the number of particles, 
Eq.~(\ref{velocityBD}) would be the asymptotic velocity
that takes into account the discreteness of $u$.

Brunet and Derrida \cite{BD} conjectured and checked that, 
within a specific numerical model,
Eq.~(\ref{velocityBD}) is actually the right answer to the stochastic problem.
They replaced the full stochastic evolution by a mean field equation
supplemented by a cutoff that simulates the discreteness in the
particle number, and that may be implemented, for example, as
\begin{equation}
u(x,t+\Delta t)=\langle u(x,t+\Delta t)|\{u(x,t)\}\rangle\,
\Theta[\langle u(x,t+\Delta t)|\{u(x,t)\}\rangle-1/N]\ ,
\label{BDtrick}
\end{equation}
where $\langle u(x,t+\Delta t)|\{u(x,t)\}\rangle$
is given by Eq.~(\ref{meantoy}).
They were able to compute the constant $c=6/\pi^2$ \cite{BD} which
turns Eq.~(\ref{velocityBD}) into Eq.~(\ref{vBD}), 
and found the profile of the front~(\ref{shapeBD}).
What Eq.~(\ref{velocityBD}) actually represents is the mean velocity
of the front when $t\gg \ln^2 N$.
The fluctuations that have been neglected so far would lead to
a random wandering of the position of the front about its mean, 
but this gives rise to subleading
effects in $t$. Indeed, according to Eq.~(\ref{sigma2Xt}),
the fluctuations of $X_t$ about its mean are proportional to
$\sqrt{t}$, which is subleading with respect to the mean 
displacement from~(\ref{velocityBD}), proportional instead to $t$.

\begin{figure}
\begin{center}
\epsfig{file=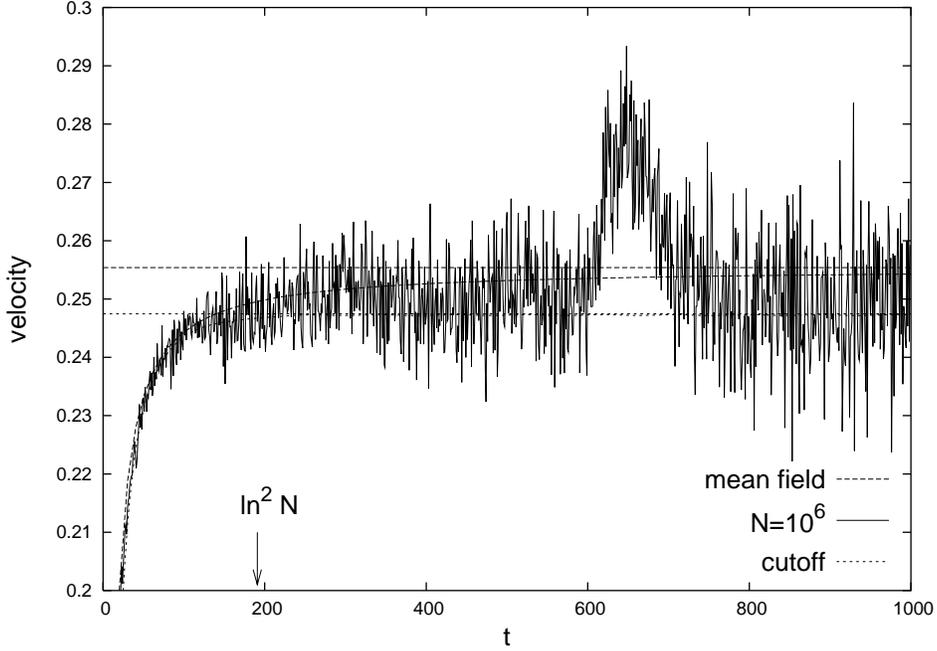,width=\wid}
\end{center}
\caption{\label{velocity2}
Instantaneous velocity of the front as a function of time (full fluctuating line)
from a numerical solution of the stochastic evolution equation~(\ref{stochatoy})
for $N=10^6$ particles per site.
Large dashed curves: analytical mean field solution, asymptotics and leading
correction. Short dashed curves: numerical solution of the mean field equation
supplemented by a cutoff  (Eq.~(\ref{BDtrick})), and analytical asymptotics
(Eq.~(\ref{vBD})) (straight dashed line).}
\end{figure}

\begin{figure}
\begin{center}
\epsfig{file=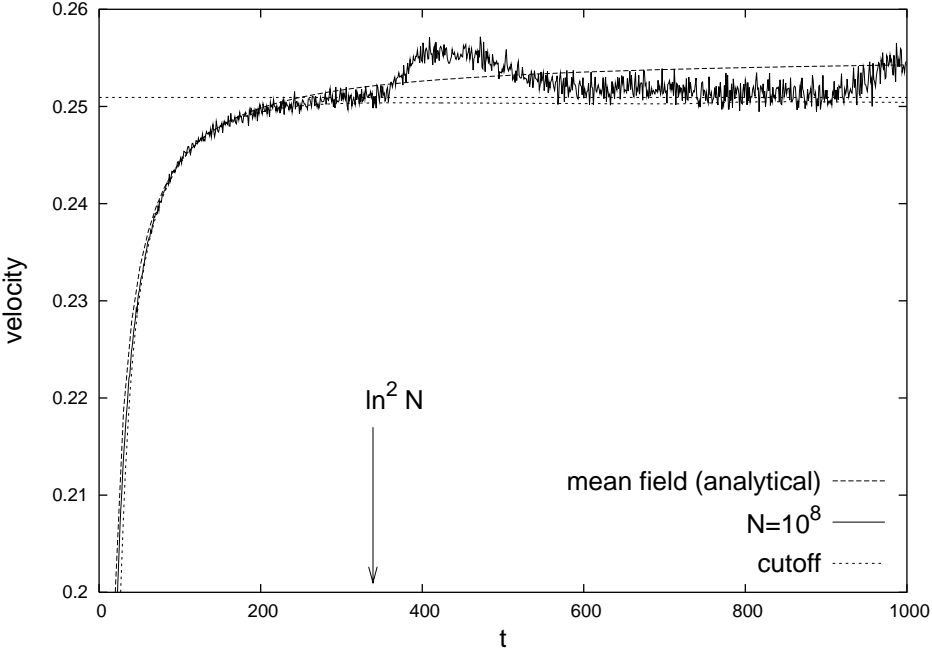,width=\wid}
\end{center}
\caption{\label{velocity3}The same, for $N=10^8$ particles per site.}
\end{figure}

In Figs.~\ref{velocity2} and~\ref{velocity3}, 
we have followed the time
evolution of the front velocity of a given realization.
The numerical calculation is
compared to the analytical result for the mean field evolution~(\ref{velocityt}), 
and the numerical
result for the modified mean field evolution~(\ref{BDtrick}).
We picked two different values of $N$ for Figs.~\ref{velocity2} and~\ref{velocity3}
respectively.
First of all,
as anticipated, up to a time of order $\ln^2 N$, the
stochastic evolution has very few fluctuations, and
follows accurately the analytical prediction~(\ref{velocityt}), 
and the numerical solution of 
Eq.~(\ref{BDtrick}).
The fact that the matching
with the leading order analytical prediction~(\ref{velocityt})
is quantitatively so good is not surprising:
indeed, the prescription that we have chosen for the measurement of the front
velocity, Eq.~(\ref{defpos2}), gives weight to the bulk of 
the front rather than to the tip,
and thus would correspond to a value of $\kappa$ of order 1 
if the definition~(\ref{defpos1})
were chosen. As demonstrated in our study of 
the BK equation in Sec.~III, for such values of $\kappa$,
the leading order analytical velocity is an accurate representation of the numerics.
As soon as we approach $t\sim\ln^2 N$, the rise of the
front velocity stops. The latter assumes a steady mean,  consistent with both the
analytical calculation (\ref{velocityBD}) and the numerical solution 
for the modified mean field equation (\ref{BDtrick}).
Note however that at this point, the velocity starts to exhibit 
large short lived fluctuations. In addition, sometimes a large fluctuation
appears, that lives a time of order $\ln^2 N$, and that drives the velocity
well above its average value expected from mean-field considerations.

\begin{figure}
\begin{center}
\epsfig{file=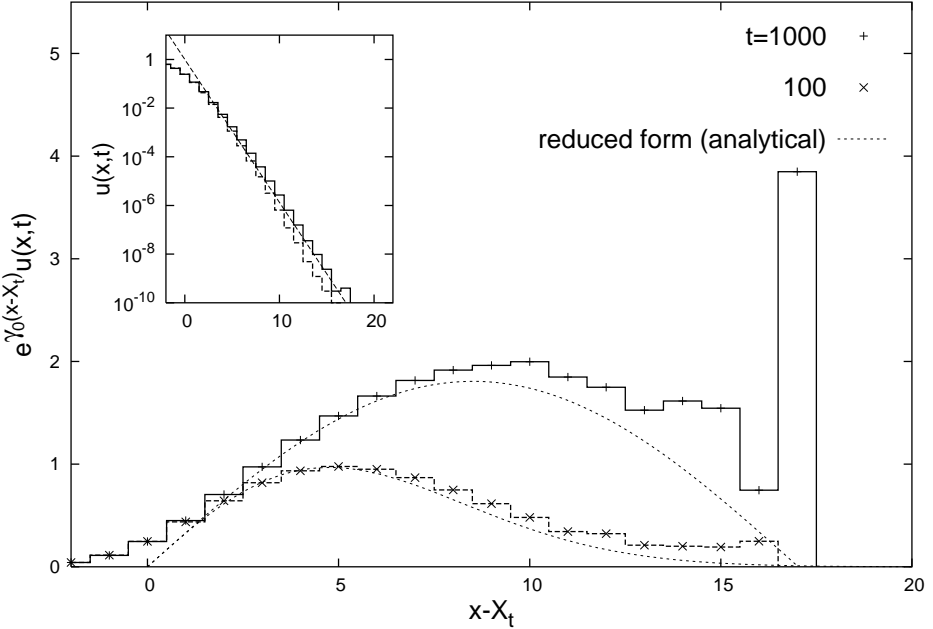,width=\wid}
\end{center}
\caption{\label{front3}Shape of the front after an evolution
of the toy model 
over $t_1=100$ (dashed steps) and $t_2=1000$ (full steps) steps of time.
{\it Insert:} The front $u(x,t)$ compared to $e^{-\gamma_0 x}$.}
\end{figure}

The deterministic building of the front until the time discreteness is
felt can be viewed also on a picture of the front at different stages of
the evolution. This is shown on Fig.~\ref{front3}.
We choose $N=10^{10}$, and 2 different times: $t_1=100$, which lies within the diffusion
time, and $t_2=1000$, which is much larger.
We plot the full front $u(x,t)$ (inset) as well as the reduced front 
$e^{\gamma_0(x-X_t)}u(x,t)$. We see that before the diffusion time,
the exponential decrease is seen in a very limited $x$-range, whereas once the
diffusion time is reached, it extends almost down to $u\sim 1/N$.
The reduced front for $t=100$ exhibits the typical Gaussian shape that
is expected from a mean field evolution, as in Eq.~(\ref{eq-redfront}). We have also 
displayed the analytical expectation.
The fact that the matching is not perfect in the large $x-X_t$ tail
is consistent with our findings of Sec.~III in the BK case.
Note that the fluctuations are small, and exclusively concentrated in
the tail of very large $x-X_t$, well ahead of the geometrical scaling zone.
For $t=1000$, the profile has changed, and looks more like an arc of sine
as predicted by Eq.~(\ref{shapeBD}).
This time, there are large fluctuations on the edge of the right part of the plot.

\begin{figure}
\begin{center}
\epsfig{file=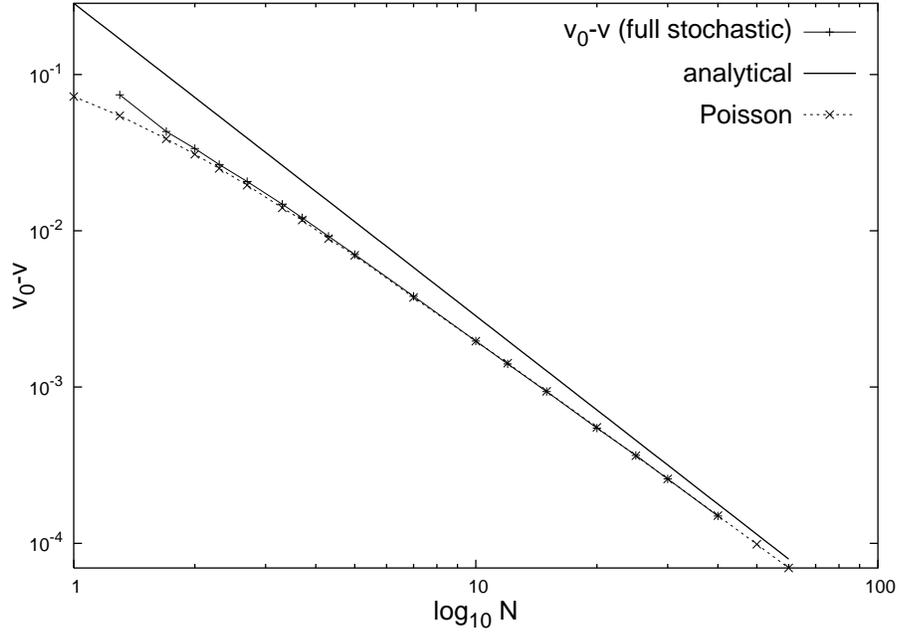,width=\wid}
\end{center}
\caption{\label{vmv0}$v_0-\langle v\rangle$ from the numerical solution
of the toy model, see Eq.~(\ref{numv0mv}) 
(full curve) compared to the analytical prediction~(\ref{vBD})
(full straight line).
The result from the simplified toy model Eq.~(\ref{poisson}) is also displayed
(dashed curve).}
\end{figure}

Finally, we check quantitatively the validity of the scaling given by 
Eqs.~(\ref{vBD}) (or~(\ref{velocityBD})) by computing numerically
$v(\gamma_0)-\langle v\rangle$. The result is displayed on Fig.~\ref{vmv0}.
Technically, we generate typically 1000 realizations of the evolution of the 
initial condition over a time $t_f\sim 10 \times t_{diff}$, and compute
\begin{equation}
\langle v\rangle=
\left\langle\frac{v(t_f)-v(t_{diff})}{t_f-t_{diff}}\right\rangle\ ,
\label{numv0mv}
\end{equation}
where the brackets on the r.h.s. stand for an average over the
realizations.
The result is displayed on Fig.~\ref{vmv0}, together 
with the analytical expectation.
We see that asymptotically, the numerical calculation approaches
slowly the analytical formula (\ref{velocityBD}) obtained from 
the mean field equation with
a cutoff
(obtained through the Brunet--Derrida procedure~(\ref{BDtrick}) 
and leading
to Eqs.~(\ref{vBD})).
Only for very large values of $N$, the agreement is actually good.
For the values of $N$ of interest for QCD, typically $N=1/\alpha_s^2\sim 50-100$, 
the analytical calculation is quite far from the numerical calculation.
However, as can be seen on the figure, the discrepancy amounts essentially
to a slowly varying function of $N$.
Note that the curve approaches the asymptote from below, which means that
the average velocity is always {\it larger} than the result from 
the mean field calculation with a cutoff.
This can be understood from our previous discussion and from 
Figs.~\ref{velocity2} and~\ref{velocity3}: the fluctuations neglected in the
Brunet-Derrida procedure pull the front ahead of its 
mean position, and the discrepancy
seen on  Fig.~\ref{vmv0} may be traced to 
the large fluctuations in the
instantaneous velocity. No more quantitative 
understanding of these deviations 
has been achieved up to now.
We have also plotted on the same
figure the result of a calculation in which
the evolution equation is mean field everywhere, except for the
foremost bin, as explained in the introduction to this section.
We see an almost perfect matching with the fully stochastic model, except for
very small values of $N$: this confirms
the observation in \cite{BD} that essentially, only the stochasticity in the
rightmost bin plays a role. However, the discrepancy at small $N$ indicates a
breaking of universality: the details of the model 
clearly start to enter.


\subsection{Variance of the position of the front}

So far, we have followed the evolution of 
one single realization. 
Each such realization undergoes a stochastic evolution given by 
Eqs.~(\ref{stochatoy})-(\ref{stocharule}).
At each time $t$, $u(x,t)$ has the universal
shape~(\ref{shapeBD}), up to fluctuations concentrated
in its tail  $u(x,t)\sim 1/N$, see Fig.~\ref{front3}.
As has been checked in Sec.~\ref{subsec_numsol},
 the average $\langle v\rangle$ exhibits
the $t$-dependence given by Eq.~(\ref{velocityBD}).
However, since the actual evolution is stochastic, the front position
$X_t$ undergoes a random walk about its mean $\langle X_t\rangle$,
and $X_t$ gets a variance $\sigma^2$. 

This fact may be understood qualitatively
in the following way. Most of the time,
the front moves forward in $x$ through a deterministic time evolution
of its bulk, which has width $\ln N/\gamma_0$. But
it may happen stochastically that several bins get successively
filled by particles much ahead of the front. Those particles will then
diffuse and multiply, and eventually form a new front that will be ahead of
the old ``deterministic'' one. The net effect is a jump in the front velocity,
one of those that can be seen on Figs.~\ref{velocity2},~\ref{velocity3}.
This phenomenon is of diffusive nature, thus we eventually expect 
the dispersion $\sigma^2$ of the positions of the front to
be proportional to $t$, as in Eq.~(\ref{sigma2Xt}).

\begin{figure}
\begin{center}
\epsfig{file=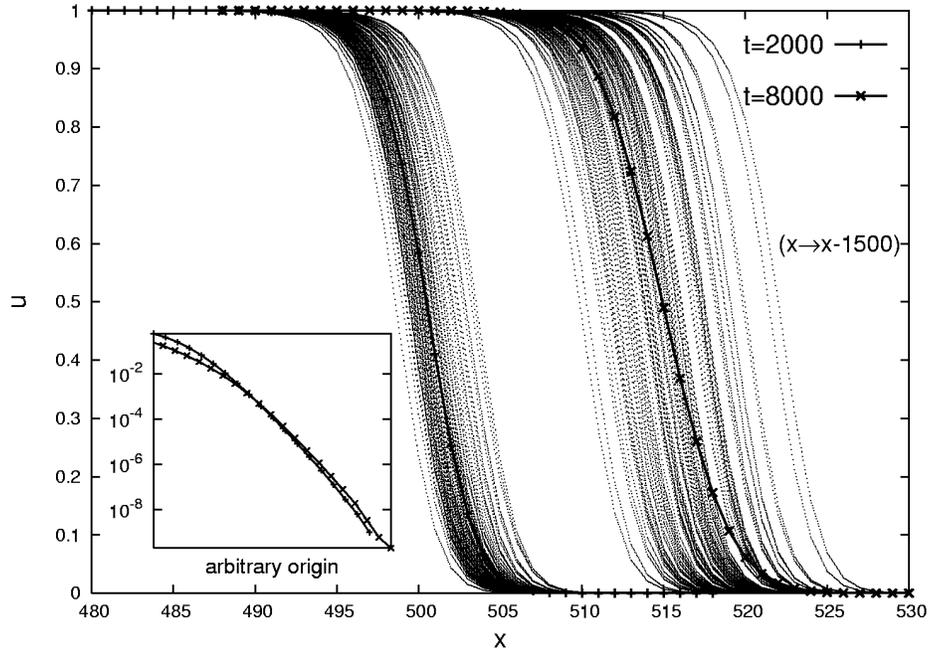,width=\wid}
\end{center}
\caption{\label{front4}1000 realizations of the evolution of the toy model
between time 0 and $t_1=2000$ (left bunch of curves), and $t_2=8000$
(right bunch of curves). {\it Insert:} the average front for these two times.}
\end{figure}

We generate 1000 realizations of the stochastic evolution of our initial
condition over the time intervals up to $t_1=2000$ and $t_2=8000$ respectively.
The dispersion of the front positions is illustrated in Fig.~\ref{front4}.
We see that there is roughly a factor of $\sqrt{t_2/t_1}=2$ in
the dispersions between the two considered times, in agreement
with Eq.~(\ref{sigma2Xt}).
Each of these realizations exhibits a universal exponential
shape~(\ref{front}) in the leading
edge region, up to small fluctuations concentrated essentially in the tip.
However, when taking the average, the curves for different times
do not superimpose anymore, as seen in the insert of Fig.~\ref{front4}.
Of course, this is due to the dispersion in the position 
of the front~(\ref{sigma2Xt}).

\begin{figure}
\begin{center}
\epsfig{file=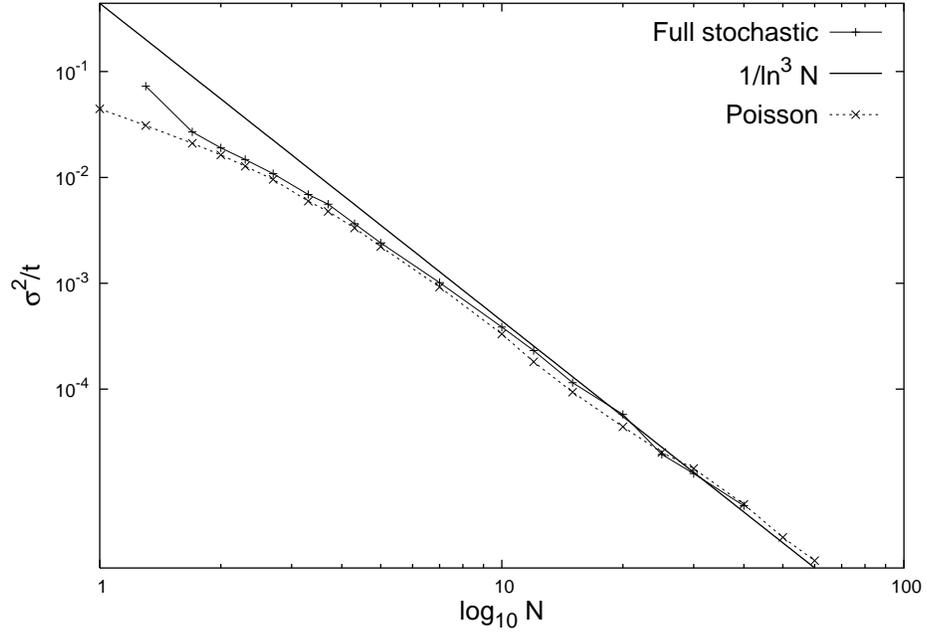,width=\wid}
\end{center}
\caption{\label{sigma}The variance of the front position divided by
the time evolution interval, see Eq.~(\ref{numsigma2}) (full curve). Line: asymptotic
prediction~(\ref{sigma2Xt}) (the overall constant being adjusted).
The result from the simplified toy model Eq.~(\ref{poisson}) is also displayed
(dashed curve).
}
\end{figure}

Now we may compute numerically the quantity
\begin{equation}
\frac{\sigma^2}{t_f\!-\!t_{diff}}=
\left\langle\frac{\left( X_{t_f}\!-\!X_{t_{diff}}
-\langle X_{t_f}\!-\!X_{t_{diff}}\rangle\right)^2}
{t_f\!-\!t_{diff}}\right\rangle
\label{numsigma2}
\end{equation}
to check the scaling form~(\ref{sigma2Xt}),~(\ref{scaling}).
The result is displayed in Fig.~\ref{sigma}.
We see again a good agreement with the expectations for
large values of $N$. Note that again, the $1/\ln^3 N$ prediction
overestimates the numerical calculation for smaller $N$.
We have displayed on the same plot 
the result of the numerical calculation in the simplified model
where stochastic effects are concentrated in the rightmost bin.
We see again that this model gives the same result as the complete
one. The discrepancies at large $N$ are statistical fluctuations,
due to the fact that the plotted curve results from an average over
a finite number of realizations (about 1000 here).
The higher moments are more and more sensitive to such fluctuations.
They should disappear when the averaging is done with more realizations.


\subsection{Consequences for QCD}

So far in this section, we have discussed exclusively the toy model.
Let us come back to the case of QCD.
The analytical results (\ref{shapeBD})-(\ref{scaling}) illustrated
on the toy model studied
in the previous section are readily taken over to QCD using 
Tab.~\ref{dictionary}. We list them here for completeness.

The scattering amplitude off a single partonic realization, 
which is not the physical
observable, but is nevertheless an important intermediate
quantity, reads
\begin{equation}
T(k,Y)\sim \frac{\ln (1/\alpha_s^2)}{\pi\gamma_0} 
\sin\left(\frac{\pi\gamma_0 \ln k^2/Q_s^2}{\ln(1/\alpha_s^2) }\right)
\left(\frac{k^2}{Q_s^2}\right)^{-\gamma_0}
\label{shapeQCD}
\end{equation}
up to fluctuations. 
Note that this formula is only valid in the window
$1\ll\ln k^2/Q_s^2\ll \ln 1/\alpha_s^2$.
The mean dependence of the saturation scale 
on rapidity reads, asymptotically 
for large rapidities
\begin{equation}
\frac{d}{dY}\langle\ln Q_s^2\rangle=\bar\alpha\frac{\chi(\gamma_0)}{\gamma_0}
-\bar\alpha\frac{\pi^2\gamma_0 \chi^{\prime\prime}(\gamma_0)}
{2\ln^2 (1/\alpha_s^2)}\ ,
\label{dqsdyQCD}
\end{equation}
and its variance
\begin{equation}
 \left\langle\left(\frac{d}{dY}\ln Q_s^2\right)^2\right\rangle
-\left\langle\frac{d}{dY}\ln Q_s^2\right\rangle^2
\propto \frac{\bar\alpha Y}{\ln^3(1/\alpha_s^2)}\ ,
\end{equation}
which, together with Eq.~(\ref{shapeQCD}), yields the asymptotic scaling form
\begin{equation}
A(k,Y)=
\langle T(k,Y)\rangle=A\left(\frac{\ln k^2-\langle \ln Q_s^2\rangle}{\sqrt{
\bar\alpha Y/\ln^3(1/\alpha_s^2)}}\right)\ .
\label{scalingQCD}
\end{equation}
This last formula quantifies the magnitude of the violations of geometric scaling.
Note that Eq.~(\ref{dqsdyQCD}) was first obtained in Ref.~\cite{MS}.
A violation of geometric scaling was also found there, but
the square root was missing in the denominator of the argument of $A$ in
Eq.~(\ref{scalingQCD}).

The formulas above are valid as long as $\bar\alpha Y\gg\ln^2(1/\alpha_s^2)$, and
asymptotically for $\alpha_s\ll 1$.
For $\bar\alpha Y\ll\ln^2(1/\alpha_s^2)$ instead, we have shown that fluctuations
do not play a role, and the
mean field discussion of Sec.~\ref{sec_meanfield} applies. This point
is particularly clear in Fig.~\ref{front3} and 
Figs.~\ref{velocity3},~\ref{velocity2}.
The kinematical regime $\bar\alpha Y\ll\ln^2(1/\alpha_s^2)$
is a window where geometric scaling should be at work.

The results obtained here are exact results of QCD. They are valid for very
small values
of $\alpha_s$, where perturbation theory is justified.
So the weak noise limit $N\gg 1$ or $\alpha_s\ll 1$ 
that we have considered here
is the consistent expansion.
Actually, as the front of the amplitude travels towards larger values of
the momenta, one should take into account the running of the coupling,
but this is just a (substantial!)
technical complication which does not change qualitatively
the picture.

Let us discuss now the validity of our results for real world QCD.
As can be seen in Figs.~\ref{vmv0} and~\ref{sigma}, the numerical 
calculations get relatively close to
the asymptotics at least up to a constant of order 1,
for say $N>10^4$, which corresponds to $\alpha_s <10^{-2}$.
Realistic attainable values of $\alpha_s$ in physics experiments are
$\alpha_s>10^{-1}$. This situation 
would rather correspond to a strong noise limit $N\sim 1$ for which,
unfortunately, no predictive theory has yet been found.


\section{Conclusion}

In this paper, we focused on 
the statistical approach to high energy QCD,
that we have developed and illustrated. We have shown that
it provides a very simple picture for high energy scattering, 
relying directly on the QCD parton model. It allows
a derivation from first principles of the most sophisticated 
analytic results that have been obtained 
so far for scattering amplitudes at fixed impact parameter. 
The simplicity of this approach
is the consequence of a deep physical connection between
the parton model and reaction-diffusion processes.
Its main interest is to gain a simple physical understanding
of the fluctuations present in the extended B-JIMWLK equations,
and to enable a clear derivation of the universal
asymptotics of the scattering amplitudes at high energy.

We have also solved numerically the QCD evolution equations
in the context of a mean field approximation, 
that leads to the BK equation.
We have confirmed that the BK equation admits traveling waves, and
studied their formation starting with different physical 
initial conditions, with different definitions of the
saturation scale. The numerical results have been compared to
the most recent analytical predictions, and a
good agreement has been found at all levels, for large enough values of the rapidity.
We have also emphasized the role of the different parts of the kernel.

To go beyond the mean field approximation to high energy scattering,
we have proposed and solved a toy model that captures the essential features of the
full QCD evolution.
The advantages of studying a toy model are numerous:
in particular, it is simple enough to allow for light and flexible 
numerical simulations.
Consequently the universal properties of the solution can be quite easily studied.
We have investigated some of these properties, confirming former numerical studies
on different models in the same universality class. We have investigated
how the asymptotics set in. One of the important conclusions of this paper is
that in the initial stages of
the evolution, the mean field solution applies.

We believe that many features would still deserve to be investigated within
such simplified models: for example, the effect of the running coupling would be crucial
both for phenomenology and for full consistency of the approach,
that relies on a small-$\alpha_s$ asymptotic expansion.
This case would correspond to reaction-diffusion in
an inhomogeneous medium, which in particular allows for
a variable maximum of particles per site.

Obviously, going beyond the leading orders in rapidity and in $\alpha_s$
will start to be model dependent (i.e. non-universal) at some order,
and the complications and specificities of QCD 
(color, 2-dimensional impact parameter space)
will show up in the evolution equations. The same is true if one wants to
give up the coarse-graining in impact parameter~\cite{MSW}.
The solutions to the more refined equations, that is
the further terms in the asymptotic expansions we have considered, will
probably be difficult to obtain from the analogy with statistical
mechanics that we have extensively used in this paper.

\vskip 1cm

\noindent
{\bf Note added:} While this work was being completed, the paper \cite{Soyez}
appeared, that also deals with numerical solutions of QCD at high energies
including fluctuations. The numerical results obtained are complementary to
ours since the attitude of the author is to focus on values of $N$ realistic
in the context of QCD, rather than emphasizing the connection to
the analytical results obtained on equations in the universality class of
the sFKPP equation.


\begin{acknowledgments}
R.E.\ is partially supported by a postdoctoral fellowship from the Swedish Research Council.
K.G.-B. acknowledges a grant of the Polish State Committee for Scientific Research
Nr. 1~P03B~028~28.
S.M.\ is a member of the Centre National de la Recherche Scientifique (CNRS), France. 
\end{acknowledgments}



\clearpage


\noindent{\bf\large Figures}

\clearpage

\end{document}